\documentclass[useAMS,usegraphicx,usenatbib]{mn2e}

\usepackage{txfonts}

\newcommand{\DERIV}[2]{\ensuremath{\frac{\partial{}#1}{\partial{}#2}}}

\newcommand{\AT}[2]{\ensuremath{\left.{#1}\right|_{#2}}}    

\newcommand{\jm}{\ensuremath{j_{\rm{m}}}}    

\newcommand{\GJ}[1]{\ensuremath{#1_{\textrm{\tiny{}GJ}}}}    

\newcommand{\RS}[1]{\ensuremath{#1_{\textrm{\tiny{}RS}}}}    
\newcommand{\RSarg}[2]{\ensuremath{#1_{\textrm{\tiny{}RS},#2}}}    

\newcommand{\PC}[1]{\ensuremath{#1_\mathrm{pc}}}    

\newcommand{\NS}[1]{\ensuremath{#1_\textrm{\tiny{}NS}}}    

\newcommand{\lambdaD}{\ensuremath{\lambda_\textrm{\tiny{}D}}}    

\newcommand{\lambdabarC}{\ensuremath{\lambdabar_\textrm{\tiny{}C}}}

\graphicspath{{./figures/}}

\title[Pair cascades in magnetospheres of NS -- I.]%
{Time-dependent pair cascades in magnetospheres of neutron stars  I.\\
  Dynamics of the polar cap cascade with no particle supply from the
  neutron star surface.}

\author[A.~N.~Timokhin]{A.~N.~Timokhin$^{1,2}$\thanks{E-mail: atimokhin@berkeley.edu}\\
  $^{1}$Astronomy Department, University of California at Berkeley,
  601 Campbell Hall, Berkeley, CA 94720, USA\\
  $^{2}$Sternberg Astronomical Institute, Universitetskij pr. 13,
  Moscow 119992, Russia}

\begin{document}

\date{Received ; in original form }

\pagerange{\pageref{firstpage}--\pageref{lastpage}} \pubyear{2010}

\maketitle

\label{firstpage}

\begin{abstract}
  I argue that the problem of electromagnetically driven
  electron-positron cascades in magnetospheres of neutron stars must
  be addressed starting from first principles.  I describe a general
  numerical algorithm for doing self-consistent kinetic simulations of
  electron-positron cascades -- wherein particle acceleration, pair
  creation and screening of the electric field are calculated
  simultaneously -- and apply it to model the
  \citet{Ruderman/Sutherland75} cascade in one dimension.  I find that
  pair creation is quite regular and quasi-periodic.  In each cycle a
  blob of ultrarelativistic electron-positron plasma is generated, it
  propagates into the magnetosphere leaving a tail of less
  relativistic plasma behind, and the next discharge occurs when this
  mildly relativistic plasma leaves the polar cap.  A short burst of
  pair formation is followed by a longer quiet phase when accelerating
  electric field is screened and no pairs are produced.  Some of
  freshly injected electron-positrons pairs get trapped in plasma
  oscillations creating a population of low energy particles.  The
  cascade easily adjusts to the current density required by the pulsar
  magnetosphere by reversing some of the low energy particles.  Each
  discharge generates a strong coherent superluminal electrostatic
  wave, what may be relevant for the problem of pulsar radioemission.
\end{abstract}


\section{Introduction}
\label{sec:introduction}

Rotation powered pulsars remain a profound puzzle despite the fact
that the first pulsar was discovered 40 years ago
\citep{HewishBell1968}.  Pulsar is a rapidly rotating strongly
magnetized neutron star (NS), as it was originally proposed by
\citet{Gold1969} and \citet{Pacini1967}, with most of its radiation
produced in the magnetosphere.  However, there is still no consistent
quantitative pulsar model.  Proposed models range from a NS with
charge starved electrosphere \citep{Krause-Polstorff1985} to a NS with
force-free magnetosphere, where acceleration of particles and, hence,
emitting zones are localized in very small spatial regions \citep{GJ}.

The force-free magnetosphere model is favored by majority of
astrophysicists working on pulsars.  There are observational hints
favoring this model: i) young pulsars produce relativistic winds with
particle number density much larger than it is necessary to screen
accelerating electric field parallel to the magnetic field; ii) pulse
peaks are narrow, what points to smallness of emitting regions, and,
hence, to smallness of regions where particles are accelerated.  From
theoretical point of view, as it was pointed out by
\citet{Sturrock71}, physical conditions in the polar cap of pulsar are
almost ideal for generation of electron-positron pair plasma.  The
energy density of the generated plasma is negligible compared to the
energy density of the magnetic field near the NS.  The magnetosphere,
if filled with plasma, almost certainly being force-free (almost
everywhere) near the NS should be force-free at much larger scales as
well; at least numerical simulations of the force-free magnetosphere
of an aligned rotator have shown that magnetosphere can remain
force-free up to distances much larger than the light cylinder radius
\citep{Timokhin2006:MNRAS1}.

Therefore, pursuing the force-free model as a 'standard model' seems
to be reasonable.  Recently the force-free pulsar magnetosphere model
has been studied in great detail
\citep[e.g.][]{CKF,Gruzinov:PSR,Timokhin2006:MNRAS1,Timokhin2007:MNRAS2,
  Spitkovsky:incl:06,Kalapotharakos2009,BaiSpitkovsky2010a}.
Force-free magnetosphere is a restricted MHD system which does not
admit any current density distribution.  By fixing the boundary
conditions -- in the case of pulsar these are the variation of
accelerating potential across the polar cap and the size of the
corotating zone -- one fixes the current density distribution.  It
turns out that the admitted range of current density distributions in
the force-free magnetosphere with realistic boundary conditions --
when the potential drop in the polar cap is less than a vacuum one --
is rather limited
\citep{Timokhin2006:MNRAS1,timokhin::PSREQ_london/2006,Timokhin2007:MNRAS2}.
For young pulsars, where potential drop in the polar cap must be
small, the current density is not constant and strongly deviates from
the Goldreich-Julian (GJ) current density $\GJ{j}\equiv\GJ{\eta}c$
($\GJ{\eta}$ is the GJ charge density, $c$ is the speed of light);
along some magnetic field lines it has the sign opposite to the sign
of the GJ charge density.

Pair production in the polar cap of pulsar is vital for sustaining of
the force-free magnetosphere -- without it there will be not enough
plasma to cancel the accelerating electric field.  Currents flowing in
the open field line zone of the magnetosphere flow through the
pair-producing region at the base of the polar cap; therefore, any
model of the polar cap cascade zone must agree with a global
magnetosphere model on the current density flowing along magnetic
field lines.  Many previously proposed models for polar cap cascades
(and almost all quantitative models) assumed stationary unidirectional
outflow of a charge separated particle beam
\citep[e.g.][]{Arons1979,Daugherty/Harding82,Muslimov/Tsygan92,
  Harding/Muslimov:heating_2::2002,Hibschman/Arons:pair_multipl::2001}.
All these models predicted current density being almost equal to the
GJ current density everywhere in the polar cap of pulsar.  This
prediction is in strong disagreement with the force-free magnetosphere
model: for young pulsars like Crab a deviation of the charge density
from $\GJ{\eta}$ of the order of few per cents -- and in
unidirectional flow this implies the same deviation of the current
density from $\GJ{j}$ -- can account for all pulsar emission.  Both
sides of this discrepancy are based on detailed simulations and it is
not possible to change some parameters in order to fit the models
together.  So, either the \emph{magnetosphere} is non-force-free or
non-stationary (or both) or \emph{polar cap cascades} do not operate
according to the existing models.

From the energetic point of view a stationary (on the rotation time
scale) force-free configuration seems to be the most preferable state
of the magnetosphere.  The inductance of the magnetosphere is much
larger than that of the polar cap, therefore, the current density in
the polar cap will be set by the magnetosphere and not in the opposite
way \citep[e.g.][]{Mestel1999}.  In my opinion these are strong hints
that existing quantitative models for particle acceleration and pair
production in pulsar polar cap do not work.  Particle acceleration and
electron-positron pair production in cascade zones can be essentially
non-stationary: time intervals of effective particle acceleration
could alternate with intervals when the accelerating electric field is
screened by electron-positron pairs created in the cascade; in fact,
in the first paper on pulsar cascades \citep{Sturrock71} the particle
flow was assumed to be non-stationary.  The current density flowing
through non-stationary cascade fluctuates strongly and the amplitude
of the fluctuation should depend on the microphysics of the
pair-generation process, not on the global physics of the
magnetosphere. However, the characteristic timescale of polar cap
cascades (microseconds) is much shorter than the magnetospheric
timescale (longer than milliseconds) so that all fluctuations due to
cascade non-stationarity will be washed out.  The average current
density in the cascade zone could be adjusted to the current density
required by the magnetosphere by adjusting the time cascade spent in
``active'' and ``passive'' phases.  On the other hand, it is still
possible that particle flow in the cascade zone is nevertheless
stationary but not unidirectional -- with some particles trapped in a
non-trivial accelerating potential \citep{Arons2009}.  However, all
these qualitative statements have to be proved.

Electromagnetically driven electron-positron cascades can operate not
only in polar caps of radiopulsars.  Some pulsars in outer parts of
their magnetospheres -- close to the place where the GJ charge density
changes the sign -- could have so-called ``outer-gap'' cascade zones
\citep{Cheng/Ruderman76}; although it seems that such acceleration
zone can exist only if polar cap cascades fail to supply enough pair
plasma to short-out the electric field in the entire magnetosphere.
Electromagnetically driven cascades should generate plasma in
magnetospheres of magnetars along open \citep{Thompson2008} as well as
closed magnetic field lines \citep{Beloborodov2007}.
Electron-positron cascades can also work in magnetospheres of black
holes \citep{Beskin1992}.  The study of pair cascade dynamics is,
therefore, of significance for a broad class of astrophysical
problems.

Non-stationary regime of electromagnetically driven cascades is poorly
investigated.  Only few attempts have been made before to construct
quantitative models for non-stationary cascades.
\citet{AlBer/Krotova:1975} were the first, their model was 0D -- it
accounted only for variability in time.  It predicted strong time
variability in pair creation rate due to the delay between emission of
a high energy photon and its decay into an electron-positron pair.
\citet{Fawley/PhDT:1978} tried to make a numerical model for
\citet{Ruderman/Sutherland75} cascade using 1D Particle-In-Cell (PIC)
code and a simple version of on-the-spot approximation for pair
injection.  At that time it was a formidable numerical problem;
simulations could be performed only for a very short time after
cascade ignition, so that no conclusive results could be drawn from
them.  \citet{Levinson05,Luo/Melrose2008,Melrose2009} used 1D
two-fluid approximation for electron-positron plasma and on-the-spot
approximation for pair injection; they studied polar cap cascades
operating in the space charge limited flow regime and found that
generation of pairs is essentially turbulent -- pair were created
throughout all physical region admitting pair creation.
\citet{Beloborodov2007} studied pair cascades in the closed field line
zone of magnetar magnetosphere.  They used on-the-spot approximation
for pair injection and tracked motion of electrons and positrons in
self-consistently calculated electric field; the electric field was
assumed to be zero at both ends of the field line.  They too concluded
that pair creation is turbulent.

In all of these models some or other simplifying assumptions about
physical processes at play were used.  It is difficult to draw
decisive conclusions about the character of particle flow pattern from
them because it is not clear a priori whether ignoring one of the
aspect of cascade physics can result in qualitatively different
behavior or not.  In my view, the study of electron-positron cascades
should be done starting ab initio.  No assumptions about the character
of particle flow should be made and the key ``ingredients'' of the
system must be preserved in the model: back reaction of particles on
the accelerating electric field and the delay between photon emission
and pair injection.  Possible complexity of system behavior compels to
conduct a numerical experiment where particle acceleration, pair
production and variation in the accelerating electric field are
modeled self-consistently.

With this paper I intend to start a series of publications dedicated
to self-consistent numerical modeling of full kinetics of
electron-positron pair cascades in magnetospheres of neutron stars.
In this paper I describe a numerical algorithm for self-consistent
modeling of electromagnetic cascades starting from first principles
and apply it for study of the most simple model of polar cap cascade
-- when particles cannot escape the NS surface -- the
\citet{Ruderman/Sutherland75} model.  The goal of this work is not
merely to quantify the Ruderman-Sutherland model but to try to infer
basic properties of electromagnetic cascades.  The most important
qualitative questions about basic cascade properties I try to answer
are i) what is the character of plasma flow and ii) how the pair
cascade adjusts to the current density required by the magnetosphere.

The structure of the paper is as follows. In
Sec.~\ref{sec:gener-numer-algorithm} I describe the general numerical
algorithm I developed for modeling of electromagnetic cascades.  In
Sec.~\ref{sec:1D-discharge} I describe physical and numerical aspects
of the of polar cap cascade model.  Simulations results and their
analysis are presented in Sec.~\ref{sec:results-numer-model}; I
summarize the inferred cascade properties in
Sec.~\ref{sec:summ-casc-prop}.  In Sec.~\ref{sec:discussion} I discuss
limitations of the model, applicability of physical approximations
used in previous works, and implication of the results for physics of
radiopulsars.

\section{General numerical algorithm}
\label{sec:gener-numer-algorithm}

\begin{figure*}
  \begin{center}
    \includegraphics[width=0.8\textwidth]{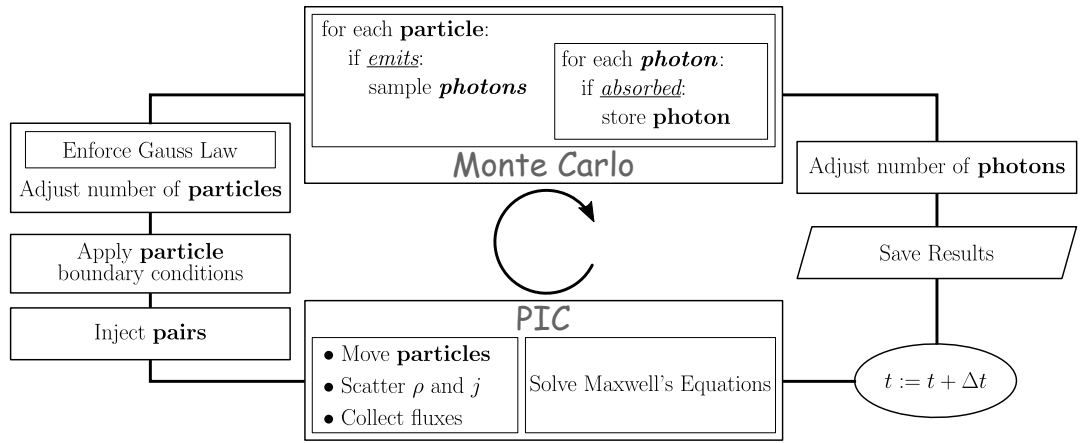}
  \end{center}
  \caption{Code structure -- sequence of operations performed at every
    time step.}
  \label{fig:code}
\end{figure*}

In electromagnetically driven pair cascade in NS magnetosphere the
following physical processes determine the behavior of the system:

\begin{enumerate}

\item Charged particles -- electrons and positrons -- are accelerated
  by the electric field induced by NS rotation.

\item Particles emit high energy gamma-rays.  The radiation mechanisms
  relevant for pulsars include curvature radiation, inverse Compton
  scattering (in both resonant and non-resonant regime) of thermal
  X-ray photons emitted by the NS, and synchrotron radiation of
  freshly created pairs \citep[e.g.][]{SturnerDermer1995,Zhang2000}.

\item Gamma photons propagate some distance and then create
  electron-positron pairs.  In pulsar polar cap the dominating process
  is single photon pair creation in the strong magnetic field
  \citep{Sturrock71}.  In the outer pulsar magnetosphere the dominant
  process will be photon-photon pair creation either on soft photons
  emitted by the NS (thermal X-rays) or soft photons produced in the
  cascade itself \citep{Cheng1986}.

\item Creation of electron-positron pairs increase plasma density and
  changes the electric field: if a pair is created in a region with
  strong electric field, electron and positron are accelerated in
  opposite directions and redistribution of the charge density alters
  the accelerating electric field.

\end{enumerate}

Probably the best numerical technique for self-consistent modeling of
plasma kinetics -- acceleration of charged particles and changes of
electromagnetic fields induced by their motion (items 1,4 in the list)
-- is Particle-In-Cell \citep[e.g.][]{Birdsall1985}.  There particle
distribution is modeled directly by representing plasma by an ensemble
of macroparticles.  PIC is a mature numerical technique, many of its
properties are well known and are subject of constant ongoing
investigations \citep[e.g.][]{Verboncoeur2005}.  Although on the
current stage of the project 1D modeling is used, PIC allows
straightforward generalization for multi-dimension.  Particle emission
and creation of electron-positrons pairs -- a radiation transfer
problem (items 2 and 3) -- in a system with strongly and rapidly
changing particle energy distribution are best to model utilizing
Monte Carlo technique \citep[e.g.][]{Sobol,Fishman1996}; the
computational costs of Monte Carlo are almost the same for 1D and
multidimensional cases.  On the other hand, in PIC plasma is already
represented by discrete particles, what makes Monte-Carlo a natural
choice.  For modeling of pair cascades I decided to develop a new
hybrid PIC/Monte Carlo code.  The existing hybrid codes used for
modeling of gas discharges do not include radiation transfer and
account only for interaction between charged and neutral particles;
Monte Carlo technique there is used to account for interaction
randomness.

The mean free path of gamma-photons does not depend on the plasma
density in the polar cap, it is set by the strength of the magnetic
field and by the curvature of magnetic field lines.  For the minimum
mean free path of photons the estimate of
\citet{Ruderman/Sutherland75} can be used, which gives
$\lambda_{\rm{mfp}}\sim10^3$~cm.  In space charge limited flow
models photon mean free path could be comparable to the NS radius,
$\lambda_{\rm{mfp}}\sim10^6$~cm.  Characteristic plasma scales are
of the order of the Debye length which depend on plasma density.  A
rough estimate for the Debye length can be made assuming plasma
density being equal to the GJ number density $\GJ{n}=\GJ{\eta}/e$:
\begin{equation}
  \label{eq:lambda_Debye}
  \lambdaD^{\textrm{\tiny{}GJ}} \sim \frac{c}{\omega_{p}^{\textrm{\tiny{}GJ}}} = 
  c \left(\frac{4\pi\GJ{\eta}e}{m_e}\right)^{-1/2}
  \simeq 2 B_{12}^{-1/2}P^{-1/2}~\mbox{cm}\,,
\end{equation}
where $B_{12}$ is the pulsar magnetic field in units of $10^{12}$~G
and $P$ is the pulsar period in seconds.  The photon mean free path is
much larger than the Debye length, and so it sets the macroscopic
scale -- the length of the computational domain.  The Debye length of
the plasma sets the microscopic scale of computations -- the cell
size.  It will be unwise to advance photons in space at the same pace
as particles -- photons propagate large distance to the absorption
point without interaction while particle motion can change on very
small spatial scales.  Propagation of photons must be calculated
separately, with larger spatial steps.

For modeling of electromagnetic cascades in NS magnetospheres I
developed a general algorithm which calculates plasma motion and
photon propagation in different numerical pace.  The scheme of the
algorithm is presented on Fig.~\ref{fig:code}, where the sequence of
operations performed at every time step is shown.

The plasma dynamics is done with the standard PIC algorithm.  Using
the current density known from the previous step I solve Maxwell
equations and get electric field at grid points.  Then for each
particle I interpolate the electric field to the particle's position
and get the electric force on the particle.  Solving the equation of
motion I advance particle momenta and positions.  Particle motion
through the cell boundary is counted as its contribution to the
electric current.  The electric current for each cell boundary is
computed simultaneously with particle motion and is stored for the
next time step.

Photon emission and pair production are calculated as follows.  I
sample how many photons capable of producing electron-positron pair
each particle emits during the current time step.  For each emitted
photon its energy is sampled from the spectral energy distribution of
the corresponding emission process.  Then I sample the distance the
photon will travel until it is absorbed.  Calculation of the optical
depth to pair creation is done with the space steps adjusted according
to the current value of the cross-section for photon absorption; most
of the steps are much larger that the cell size.  In this way photon
propagation is done in the (appropriate) and much faster numerical
pace than particle advance.  Photon's energy, position and time of
absorption are stored in an array.  At every time step I iterate over
the photon array and pick up photons which are absorbed at the current
time step.  For each of the selected photons I inject an electron and
a positron at the point of photon absorption and delete that photon
from the array.  Being injected at the same point freshly created
electron and positron do not contribute to the charge and current
densities at the time step of injection.

If there are too many particles of particular kind in the
computational domain, their number can be reduced by deleting some
randomly selected particles.  The total statistical weight of the
selected particles is stored and then statistical weights of all
remaining particles of the same kind are increased in order to
compensate for the deleted particles.  Although this conserves the
overall charge of the system, the resulting charge distribution will
be slightly different from the one before particle deletion.  To
proceed with charge conserving algorithm one need to solve Poisson
equation in order to bring the electric field in accordance with the
altered charge distribution.  When the number of photons is reduced,
the later step is, of course, unnecessary.

\section{One-dimensional discharge in pulsar polar cap}
\label{sec:1D-discharge}

As previously there were no truly self-consistent studies of
electromagnetic cascades -- allowing time-dependence and incorporating
all classes of relevant microscopic processes -- I decided to address
first the simplest case in order to develop an intuition about physics
of pair plasma generation.  It was not clear a priori what is the
pattern of plasma flow, and in order to develop an appropriate
numerical technique to model a realistic system with many microscopic
processes at play a ``bare-bone'' model must be studied first.

\subsection{Physical model}
\label{sec:physical-model}

The \citet{Ruderman/Sutherland75} model for pair cascade in the polar
cap of pulsar is the simplest possible model for a pair cascade.
Ruderman and Sutherland considered the case when the NS angular
velocity is anti-parallel to its magnetic momentum -- so that the
Goldreich-Julian charge density is positive -- and assumed that the
work function to extract a positive ion from the surface of a NS is
much larger than the available electric potential.  In this model
there is no plasma inflow from the surface of the NS and all plasma in
the cascade zone is produced by pair creation in a series of
'discharges'.  When enough plasma is produced in the discharge zone,
it screens the accelerating electric field and, therefore, stops
particle acceleration and pair creation.  Plasma flows into the
magnetosphere and -- as there is no source of plasma than (now
suppressed) pair creation -- the plasma density decreases.  When there
are not enough charged particles to screen the accelerating electric
field the pair formation starts again.

Although there are hints \citep[e.g.][]{Medin2007} that the work
function in the NS crust can be small enough -- so that particles can
be extracted from the star -- and cascade operates in the so-called
space-charge limited flow regime \citep{Arons1979}, I think that
studying of the Ruderman-Sutherland model is worth the effort.  As
reasons for this I name: i) this model is intrinsically non-stationary
and should be a good test of whether non-stationary cascade is indeed
flexible enough to adjust itself to any current density required by
the magnetosphere; ii) being the simplest model it could be used as a
testing ground for the numerical technique; iii) boundary conditions
in this model (no plasma inflow) are similar to that in the problem of
electron-positron pair plasma generation near the horizon of a black
hole \citep{Beskin1992,Hirotani1998}, and, hence, from solution of the
pulsar problem it would be possible to get some hints to how to
address the latter problem.

\citet{Ruderman/Sutherland75} estimate the height of the cascade zone
for young pulsars (their eq.~(22)) as
\begin{equation}
  \label{eq:h_RS}
  \RS{h} \sim 5\times10^3 \rho_6^{2/7} P^{3/7} B_{12}^{-4/7}~\mbox{cm}\,,
\end{equation}
where $B_{12}$ is the pulsar magnetic field in units of $10^{12}$~G,
$\rho_6$ is the radius of magnetic field line curvature normalized to
$10^6$~cm and $P$ is the pulsar period in seconds.  For young pulsars,
with the period of the order of $\sim{}0.1$~sec, $\RS{h}$ is less that
the width of the polar cap
\begin{equation}
  \label{eq:r_pc}
  \PC{r}\simeq{}1.45\times{}10^4P^{-1/2}~\mbox{cm}\,.
\end{equation}
Therefore, 1D approximation should work well for such cascades.  In
the Ruderman-Sutherland model the charge density deviate strongly from
the GJ charge density what creates accelerating electric field
comparable to the vacuum electric field.  The general relativistic
effects introduce corrections to the electric field of the order of
several per cents of the vacuum electric field
\citep{Beskin1990,MuslimovTsygan1990}, and for this problem they can
be ignored.

For young pulsars the dominant emission process in terms of number of
pair-production capable photons is the curvature radiation
\citep[e.g.][]{Hibschman/Arons:pair_multipl::2001}.  In this paper I
am primarily interested in dynamics of the discharge zone, the region
with the accelerating electric field.  The size of that zone should be
of the order of few $\RS{h}$, which is of the order of the mean free
path of curvature photons.  Synchrotron photons emitted by the
injected pairs are much less energetic than curvature photons and,
therefore, the mean free path of synchrotron photons is much larger
than $\RS{h}$.  They are absorbed at large distances from the NS where
plasma density is expected to be very high and electric field is
already screened.  Hence, the pairs produced by the synchrotron
photons do not influence the discharge dynamics and synchrotron
emission can be ignored.

So, the minimal physical model for the Ruderman-Sutherland cascade
includes 1D electrodynamics, curvature radiation as the photon
emission process, and pair creation in a strong magnetic field as the
source of electron-positron pairs.

\subsection{Main equations}
\label{sec:main-equations}

In the superstrong magnetic field of pulsar charged particles are in
the first Landau level and move strictly along magnetic field lines.
The radius of curvature of magnetic field lines $\rho$ is much larger
than the polar cap radius $\PC{r}$; for distances comparable to the
width of the polar cap particle dynamics can be considered as a motion
along straight lines.  The curvature of the field lines is essential
for photon emission and pair creation.  The radius of curvature of
magnetic field lines enters in the expressions for curvature radiation
and gamma-ray absorption cross-sections as a parameter, which can
depend on particle position.

I assume that charged particles moves along straight magnetic field
lines which are perpendicular to the NS surface.  A coordinate axis
$x$ is directed along the field lines, its origin is at the NS surface
and positive direction is toward the magnetosphere.  In the
one-dimensional model charged particles are represented by thin sheath
with infinite extend in the direction perpendicular to the $x$-axis.
I normalize particle momentum to $m_ec$ -- the normalized particle
momentum $p$ is its 4-velocity $p=\beta\gamma$, where $\beta=v/c$ is
particle velocity normalized to the speed of light and
$\gamma=(1-\beta^2)^{-1/2}$ is the Lorentz factor.  The equation of
motion for a particle $i$ is
\begin{eqnarray}
  \label{eq:dx/dt}
  \frac{d x_i}{d t} & = & v_i \\
  \label{eq:dp/dt}
  \frac{d p_i}{d t} & = & \frac{e}{m_ec}\frac{\tilde{q}_i}{\tilde{m}_i} E - W_{\rm{rr}}
\end{eqnarray}
$\tilde{q}_i$ and $\tilde{m}_i$ are particle charge and mass in units
of electron charge $e$ and mass $m_e$ correspondingly.  $W_{\rm{rr}}$
is the term responsible for radiation reaction.  For curvature
radiation it is given by
\begin{equation}
  \label{eq:W_rr}
  W_{\rm{rr}}=  \frac{2e^2}{3m_ec} \frac{\tilde{q}_i^2}{\tilde{m}_i} \frac{p^4}{\rho^2}
\end{equation}
For low energy particles radiation reaction becomes very small and for
them $W_{\rm{rr}}$ in eq.~(\ref{eq:dp/dt}) is ignored (see
Sec.~\ref{sec:numer-impl}).

In one-dimensional model the only changing component of
electromagnetic fields is the electric field component $E$ parallel to
the $x$-axis.  The system is essentially electrostatic and the
electric field can be obtained from the solution of the Poisson
equation for the electric potential $\phi$
\begin{equation}
  \label{eq:d2Phi_dx2}
  \frac{d^2\phi}{dx^2} = -4\pi(\eta-\GJ{\eta})
\end{equation}
as $E=-d\phi/dx$.  Here $\eta$ is the charge density and $\GJ{\eta}$
is the GJ charge density.  In order to solve eq.~(\ref{eq:d2Phi_dx2})
one has to specify either the potential difference across the domain
or one has to set the electric field to some fixed value at one end of
the domain, either on the NS surface or at the base of the
magnetosphere; these boundary conditions must be specified at every
time step.  The main free parameter in the problem is the average
current density which flow trough the cascade zone.  Charge can
accumulate on the NS surface and particles can be send back from the
magnetosphere.  Hence, boundary conditions are different at every time
step -- the potential drop along the computation domain as well as the
electric field at the domain boundaries change with time.  Boundary
conditions are related in some complicated way to the requirement of
providing a certain value for the average current density.  However,
if charge conservation is taken into account, the electrostatic
\emph{boundary value} problem can be transformed into an \emph{initial
  value} problem, which does not require boundary conditions.  I do so
in Appendix~\ref{sec:app_1D_electrodynamics}, where I derive the
equation for the electric field%
\footnote{\citet{Levinson05} used the same approach for calculation of
  the electric field.  They did not elaborate on the physical meaning
  of $\jm$, so I decided to present a detailed derivation of
  eq.~(\ref{eq:dE_dt})}
\begin{equation}
  \label{eq:dE_dt}
  \frac{d E(x,t)}{d t} = -4\pi\left( j(x,t)-\jm(t) \right)\,.
\end{equation}
Here $j$ is the current density at point $x$ at time $t$, and $\jm$ is
the average current which flow through the calculation domain,
determined by the twist of the magnetic field imposed by the global
stress balance of the magnetosphere.  To solve this equation only an
initial configuration of the electric field in the domain
$\AT{E(x)}{t=0}$ is necessary; the boundary conditions are
incorporated in $\jm$, i.e. the electric field at domain boundaries
will adjust itself to provide the required average current density
$\jm$.

The solution of the eq.~(\ref{eq:dE_dt}) gives the correct electric
field -- the one which satisfies the Maxwell equations (in the 1D case
it is the Gauss law) -- if one starts from a configuration where $E$
is obtained as a solution of the Poisson equation and numerical
algorithm conserves electric charge.  In my simulations, at the very
first time step I set some boundary conditions on the electric field
(or the potential drop in the domain) and some initial particle
distribution; then I compute the charge density $\eta$ and solve the
Poisson equation~(\ref{eq:d2Phi_dx2}) for that boundary conditions to
get the initial distribution of the electric field in the domain.  At
all subsequent time steps for each point in the numerical grid I
compute the electric field from its value at the previous time step
using equation~(\ref{eq:dE_dt}); the current density $j$ due to
particle motion is calculated using a change conserving algorithm.

Physically, in order for electric field to be zero in the polar cap
the charge density must be equal to the GJ charge density and the
current density equal to the $j_m$ required by the magnetosphere.  The
GJ charge density enters in the Poisson equation which is solved at
the first time step; because of charge conservation the system ``keeps
memory'' of the GJ charge density at all subsequent time steps.  The
current density $\jm$ enters in the equation for the electric field.
So, the system tries to adjust to both these requirements.


Particles moving along curved magnetic field lines emit photons via
curvature radiation mechanism.  Spectral energy distribution of
curvature photons emitted by a particle with the Lorentz factor
$\gamma$ is given by the standard formula \citep{Jackson1975}
\begin{equation}
  \label{eq:dN_phcr_dt}
  \frac{\partial{}N_{\rm{ph}}}{\partial{}t\,\partial{}\epsilon}(\epsilon) =
  \frac{1}{\sqrt{3}\pi}\frac{\alpha_f\:c}{\lambdabarC}
  \frac{1}{\gamma^2} \int_{\epsilon/\epsilon_\textrm{\tiny{}CR}^{\rm{}peak}}^\infty d\xi\, K_{5/3}(\xi)\,,
\end{equation}
where $\epsilon$ is the photon energy normalized to $m_ec^2$,
$\alpha_f$ is the fine structure constant,
$\lambdabarC=\hbar/m_ec=3.86\times10^{-11}$~cm is the reduced Compton
wavelength, $K_{5/3}$ is the modified Bessel function of order $5/3$;
$ \epsilon_\textrm{\tiny{}CR}^{\rm{}peak}=(3/2)\lambdabarC\,\rho^{-1}\gamma^3 \simeq 57.92\:\rho_6^{-1}\gamma_6^3$
is the peak energy of curvature photons, $\gamma_6\equiv\gamma/10^6$.
The integral in eq.~(\ref{eq:dN_phcr_dt}) has asymptotic forms
\begin{equation}
  \label{eq:int_K_asympt}
  \int_{y}^\infty d\xi\, K_{5/3}(\xi) \simeq
  \left\{
    \begin{array}{ll}
      2.15 y^{-2/3} - 1.81,& \mbox{if }  y\ll1\\
      1.25 e^{-y} y^{-1/2},& \mbox{if }  y\gg1
    \end{array}
  \right.
\end{equation}
The total number of curvature photons with energies greater that some
$\epsilon_{a}$ emitted by the particle during time $dt$ is
\begin{equation}
  \label{eq:N_CR_emitted}
  dN_{\rm{ph}}(\epsilon>\epsilon_{a})=  dt\, \frac{1}{\sqrt{3}\pi}\frac{\alpha_fc}{\lambdabarC}
  \frac{1}{\gamma^2} F\left(\frac{\epsilon}{\epsilon_{a}}\right)
\end{equation}
where
\begin{equation}
  \label{eq:F_zeta}
  F(\zeta) = \int_{\zeta}^\infty d\xi \int_{\xi}^\infty dx\, K_{5/3}(x)\,.
\end{equation}
For small values of its argument $F(\zeta)$ has the following
asymptotic form
\begin{equation}
  \label{eq:F_zeta__small_zeta}
  F(\zeta) \simeq
  1 + 0.346 \zeta - \zeta^{1/3}(1.232 + 0.033 \zeta^2),
  \quad   \zeta\ll1\,.
\end{equation}
Only very high energy photons capable to produce an electron-positron
pair in the calculation domain are of relevance for the considered
problem and only those are tracked in the code (see
Sec.~\ref{sec:numer-impl}).


I assume that photons are emitted tangentially to the magnetic field
lines and then move along straight lines.  The angle $\psi$ between
the photon momentum and the magnetic field increases as the photon
propagates further from the emission point.  In a simple model where
magnetic field lines have constant curvature the angle between the
photon momentum and the magnetic field is given by
\begin{equation}
  \label{eq:psi_rho6}
  \psi(x) = (x-x_e)/\rho\,,
\end{equation}
where $x_e$ is the coordinate of the emission point. In the dipolar
magnetic field the expression for $\psi(x)$ is slightly more
complicated
\begin{equation}
  \label{eq:psi_dipole}
  \psi(x) \simeq \frac{3}{4}\theta_e\frac{x-x_e}{x}\sqrt{1+\frac{x_e}{\NS{R}}}\,,
\end{equation}
here $\theta_e$ is the colatitude of the emission point (a free
parameter in the 1D model), $\NS{R}$ is the NS radius.  Both models
were used in the simulations.  The cross-section of photon absorption
is given by \citep{Erber1966}
\begin{equation}
  \label{eq:sigma_BG}
  \sigma_{B\gamma}=0.23
  \frac{\alpha_f}{\lambdabarC}\frac{B}{B_q}\sin\psi\;
  \exp\left( -\frac{8}{3\chi}  \right)\,,
\end{equation}
where
$\chi=\epsilon{}B\sin\psi/B_q\simeq{}2.27\times10^{-2}\epsilon{}B_{12}\psi$,
and $B_q=m_e^2c^3/e\hbar\simeq{}4.41\times10^{13}$~G is the critical
magnetic field strength.  The cross-section grows exponentially as
photon propagates further from the emission point.


When the photon is absorbed I assume that its energy is equally
divided between newly created electron and positron.  The
perpendicular component of particle's momentum will be rapidly
radiated as synchrotron photons, which, as described before, are
neglected.  Injected particle ends up having only the longitudinal
component of the momentum
\begin{equation}
  \label{eq:p_injected}
  p_{e^{\pm}}\simeq\left(\frac{\epsilon^2-4}{4+\psi_a^2\epsilon^2}\right)^{1/2}\,,
\end{equation}
where $\psi_a$ is the angle between the photon momentum and the
magnetic field at the absorption point.

\subsection{Numerical implementation}
\label{sec:numer-impl}

In this section I describe normalization of physical quantities,
introduce several numerical parameters controlling the algorithms, and
give their typical values in my simulations. I also give an overview
of main numerical algorithms used in the code; a detailed description
of the numerical code will be given elsewhere.

Distances are normalized to the radius of the pulsar polar cap,
$x_0=\PC{r}$, given by eq.~(\ref{eq:r_pc}).  The electric potential is
normalized to the vacuum potential drop between the rotation axis and
the edge of the polar cap in the aligned rotator
\begin{equation}
  \label{eq:Phi_0}
  \Phi_0 = \frac{\Omega}{c}\frac{B\PC{r}^2}{2}
  \simeq6.6\times10^{12} B_{12}P^{-2}~\mbox{V}\,,
\end{equation}
$\Omega$ is the pulsar angular velocity.  The electric field is
normalized to
\begin{equation}
  \label{eq:E_0}
  E_0=\frac{\Phi_0}{x_0}\simeq4.6\times10^{8} B_{12}P^{-3/2}~\mbox{V/cm}\,,
\end{equation}
and the charge density to the absolute value of the Goldreich-Julian
charge density
\begin{equation}
  \label{eq:Rho_GJ}
  \eta_0\equiv\frac{\Omega{}B}{2\pi{}c}=\frac{\Phi_0}{\pi{}x_0^2}
\end{equation}


Each numerical particle is a macroparticle representing a (large)
number of real particles $N_0$, either electrons or positrons.  Each
numerical particle has a statistical weight $w_i=\tilde{w}_iN_0$.
When a macroparticle emits a photon, the latter gets the particle's
statistical weight (also see below the description of photon
sampling); when the photon is absorbed the injected electron and
positron get the photon's statistical weight.  An important numerical
parameter is $\GJ{N^{\rm{cell}}}$ -- the number of macroparticles with
the normalized statistical weight $\tilde{w}_i=1$ in a cell which
create Goldreich-Julian charge density
\begin{equation}
  \label{eq:N_0__N_GJ}
  \eta_0=eN_0\GJ{N^{\rm{cell}}}\,.
\end{equation}
The parameter controlling the number of numerical particles in the
simulation is $\GJ{N^{\rm{cell}}}$; $N_0$ is computed at the start of
the simulation from eq.~(\ref{eq:N_0__N_GJ}).  The difference in the
number density between particles of opposite signs of the order of
$\GJ{N^{\rm{cell}}}$ results in a large electric field; this number
should be not very small, otherwise numerical noise will strongly
contaminate results.  In my simulations values of
$\GJ{N^{\rm{cell}}}\ga5$ provide acceptable level of numerical noise
which allows to recognize plasma oscillation excited in the pair
plasma.

The calculation domain is divided in $M_x$ equal numerical cells; a
typical value of $M_x$ in my simulations is several thousands.  I use
1D version of the charge conservative algorithm proposed by
\citet{VillasenorBuneman92} for scattering of charge and current
densities to the grid points and for interpolation of the electric
field to particle' locations.  Integration in time of
eq.~(\ref{eq:dx/dt}),~(\ref{eq:dp/dt}),~(\ref{eq:dE_dt}) is done with
a leap-frog scheme with a uniform time step $\Delta{}t$.  The
radiation reaction in eq.~(\ref{eq:dp/dt}) is taken into account only
for particles with momentum larger than a certain value
$p_{\rm{rr}}^{\min}$.


For each particle if its momentum is larger than a certain value
$p_{\rm{rad}}^{\min}$ I calculate the mean number of photons
$N_{\rm{ph}}$ with energies larger than a certain
$\epsilon_{\rm{em}}^{\min}$ the particle emits during time $\Delta{}t$
according to eq.~(\ref{eq:N_CR_emitted}).  If $N_{\rm{ph}}$ is small,
less than a certain $N^{\max}_{\rm{ph}}$, I sample the number of
actually emitted photons from the uniform distribution with the mean
value equal to $N_{\rm{ph}}$.  Then for each photon I sample its
energy from the distribution given by eq.~(\ref{eq:F_zeta}) using
either cutpoint or inverse transform methods \citep{Fishman1996}.  The
values of $F(\zeta)$ in eq.~(\ref{eq:F_zeta}) are tabulated for
$0.01\le\zeta\le10$ for use in cutpoint method, for smaller $\zeta$
invers transform method is used with the asymptotic formula
$F(\zeta)\simeq{}1-1.232\zeta^{1/3}$.  If
$N_{\rm{ph}}>N^{\max}_{\rm{ph}}$, the particle emits a fixed number of
numerical photons -- the spectrum is divided in
$N^{\rm{bin}}_{\textrm{\tiny{}CR}}$ bins, the number of emitted
numerical photons is equal to $N^{\rm{bin}}_{\textrm{\tiny{}CR}}$;
each photon gets a statistical weight equal to the product of the
statistical weight of the emitting particle and the number of photons
emitted in the corresponding spectral energy bin given by
eq.~(\ref{eq:dN_phcr_dt}).


To calculate the position where the photon is absorbed I sample the
optical depth the phonon should achieve before being absorbed; then I
integrate the cross-section~(\ref{eq:sigma_BG}) along phonon's
trajectory until the required optical depth is reached.  The
cross-section of photon absorption in the polar cap grows
exponentially with the distance from the emission point, and most of
the trajectory do not make significant contribution to the optical
depth.  At first optical depth along photon's trajectory is calculated
using rectangle methods with large spatial steps ($\sim1/20-1/40$ of
the domain size) until the optical depth on the next step would exceed
the required value.  This integration method overestimates the optical
depth, the trajectory always continues beyond this intermediate stop
point.  I redo the cross-section integration between the emission and
the stop points using 15 points Gauss-Kronrod integrator what provides
a very accurate value for the optical depth at the stop point.  Then
the optical depth integration is proceeded with a smaller spatial step
comparable to the cell size.  The number of cross-section evaluation
in this algorithm is, on average, by a factor of few tens smaller than
a cross-section integration with the step equal to the cell size would
require.

Values of the numerical parameters $p_{\rm{rr}}^{\min}$,
$p_{\rm{rad}}^{\min}$, $\epsilon_{\rm{em}}^{\min}$,
$N^{\max}_{\rm{ph}}$, $N^{\rm{bin}}_{\textrm{\tiny{}CR}}$ are fixed at
the start of the simulation.  These values are chosen to sample all
pair creation capable photons and correctly account for the radiation
reaction on one side, and to minimize the computation time of the
other side; their particular values depend on physical conditions: the
pulsar period, the magnetic field strength, and the radius of
curvature of magnetic field lines.  The typical values I used in the
simulations:
$p_{\rm{rr}}^{\min}\sim{}p_{\rm{rad}}^{\min}\sim10^5-10^6$,
$\epsilon_{\rm{em}}^{\min}\sim2-40$,
$N^{\max}_{\rm{ph}}\sim{}N^{\rm{bin}}_{\textrm{\tiny{}CR}}\sim{}50$.


The numerical code was developed from scratch and written in C++
programming language.  Its modular object-oriented structure is
designed to facilitate further extension to multidimension and
incorporation of additional physical processes.  I tested the PIC part
of the code performing simulations of the following test problems:
oscillation of two test particles, two stream instability in both
relativistic and non-relativistic regime, non-relativistic and
relativistic Child's laws, dependence of plasma frequency on numerical
resolution \citep{Birdsall1985}.  I also tested that the code is
indeed change conservative up to machine precision.  The Monte-Carlo
part of the code was tested as follows.  I verified that the energy
distribution of emitted photons agrees with the spectrum of curvature
radiation.  For several fixed emission points, different values of
photon energy, magnetic field strength, and radius of curvature of
magnetic field lines I compared the distribution of photon absorption
points produced by the Monte-Carlo code with the corresponding
theoretical distributions.  I also checked that for a given time
interval the total energy of emitted photons is equal to the particle
radiation reaction losses.

\section{Results of numerical modeling}
\label{sec:results-numer-model}

The numerical simulations have shown that in the Ruderman-Sutherland
model pair creation is quasi-periodic and self-sustained.  I performed
simulations for different initial particle distributions, initial
electric fields, strengths of the magnetic field, radii of curvature
of magnetic field lines, and pulsar periods.  Independent on the
initial configuration for non-zero $\jm$ pair-creation process begins
some time after the start of simulations.  How and how much plasma is
formed in this initial burst depends on the specific setup.  Plasma
generation stops after enough plasma is produced to screen the
electric field.  After plasma generated in this burst of pair
formation leaves the domain (in a couple of domain flyby times),
behavior of the cascade for given magnetic field, pulsar period, and
the mean current $\jm$ is the same, independent on the initial
configuration.  All subsequent bursts of pair formation do not depend
on the initial setup -- the system seems to forget the initial
conditions.  After that initial burst of pair creation the cascade
zone always settles down to a quasi-periodic behavior.  For a given
$\jm$ cascade behavior is qualitatively similar for all other physical
parameters admitting pair creation.

I describe here main properties of the Ruderman-Sutherland cascade
using as an example a pulsar with period $P=0.2$~s, magnetic field
$B=10^{12}$~G, and the radius of curvature of magnetic field lines
$\rho=10^6$~cm.  The radius of curvature of magnetic field lines
comparable to the NS radius implies that there is a non-dipolar
component of the magnetic field in the polar cap region.  I performed
simulations for pure dipole magnetic field with $\rho\sim10^8$~cm too.
Qualitatively results do not depend on the radius of curvature, but
for smaller $\rho$ calculations with the same numerical resolution can
be done faster because the size of the gap with accelerating electric
field is smaller.  On the other hand, adoption of this value for
$\rho$ simplifies comparison with the original
\citet{Ruderman/Sutherland75} model, where the same value for $\rho$
was used.

The polar cap radius for such pulsar is $\PC{r}=3.24\times10^4$~cm
(eq.~(\ref{eq:r_pc})).  The heights of the gap should be (see
eq.~(\ref{eq:h_RS}))
\begin{equation}
  \label{eq:h_RS_num}
  \RSarg{h}{1}\simeq{}2.5\times10^3\mbox{cm}\simeq{}0.077\PC{r}\,,
\end{equation}
and the potential drop in the gap is
\begin{equation}
  \label{eq:dV_RS_num}
  \RSarg{\Delta{}V}{1}=2\pi\GJ{\eta}\RSarg{h}{1}^2\simeq{}1.98\times10^{12}\mbox{V}=0.012\Phi_0\,,
\end{equation} 
so the maximum Lorentz factor of electrons and positrons is
\begin{equation}
  \label{eq:gamma_max_RS_num}
  \RSarg{\gamma}{1}^{\max}=\frac{e\RSarg{\Delta{}V}{1}}{m_ec^2}\simeq{}3.87\times10^6\,.
\end{equation}

The angular velocity of NS rotation is anti-parallel to the magnetic
moment of the star and the Goldreich-Julian change density is
positive.  The length of the computation domain for the simulations
described in this section is $L=0.3\:\PC{r}\simeq9.72\times10^3$~cm.
Numerical grid has $M_x=5000$ points, so that the cell size
$\Delta{}x\simeq1.94$~cm.  The number of numerical particles in cell
providing the GJ charge density $\GJ{N^{\rm{cell}}}=10$.  Other
numerical parameters are
$p_{\rm{rr}}^{\min}=p_{\rm{rad}}^{\min}=5\times10^5$,
$\epsilon_{\rm{em}}^{\min}=20$, $N^{\max}_{\rm{ph}}=50$,
$N^{\rm{bin}}_{\textrm{\tiny{}CR}}=80$.

I describe properties of cascade with physical parameters given above
for 3 different current densities: $\jm=\GJ{j}$, $\jm=0.5\GJ{j}$, and
$\jm=1.5\GJ{j}$.  First I describe main properties of cascade with
$\jm=\GJ{j}$.  Pair formation dynamics for different current densities
is qualitatively similar.  Later in this section I will highlight the
differences in cascade properties for $\jm=0.5\GJ{j}$ and
$\jm=1.5\GJ{j}$.

\subsection{Pattern of plasma flow}
\label{sec:pattern-plasma-flow}

\begin{figure*}
\begin{center}
  \includegraphics[width=\textwidth]{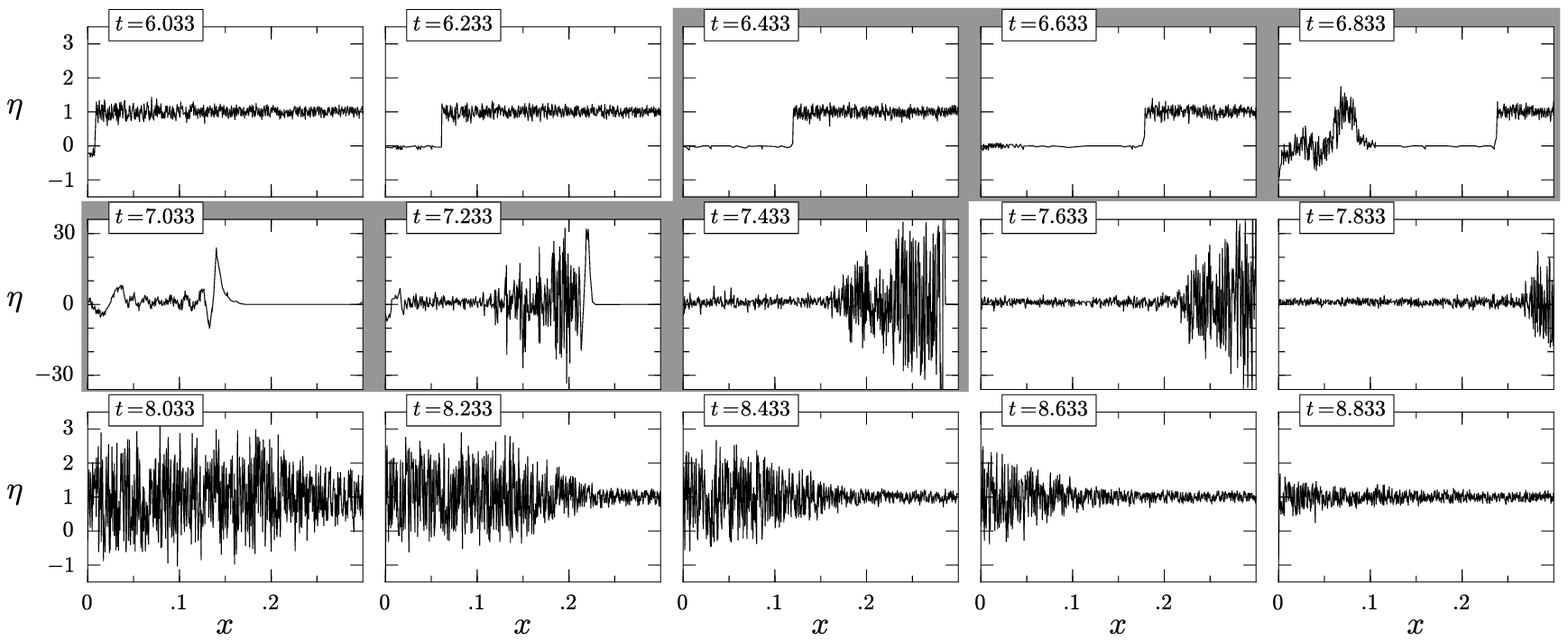}
  \includegraphics[width=\textwidth]{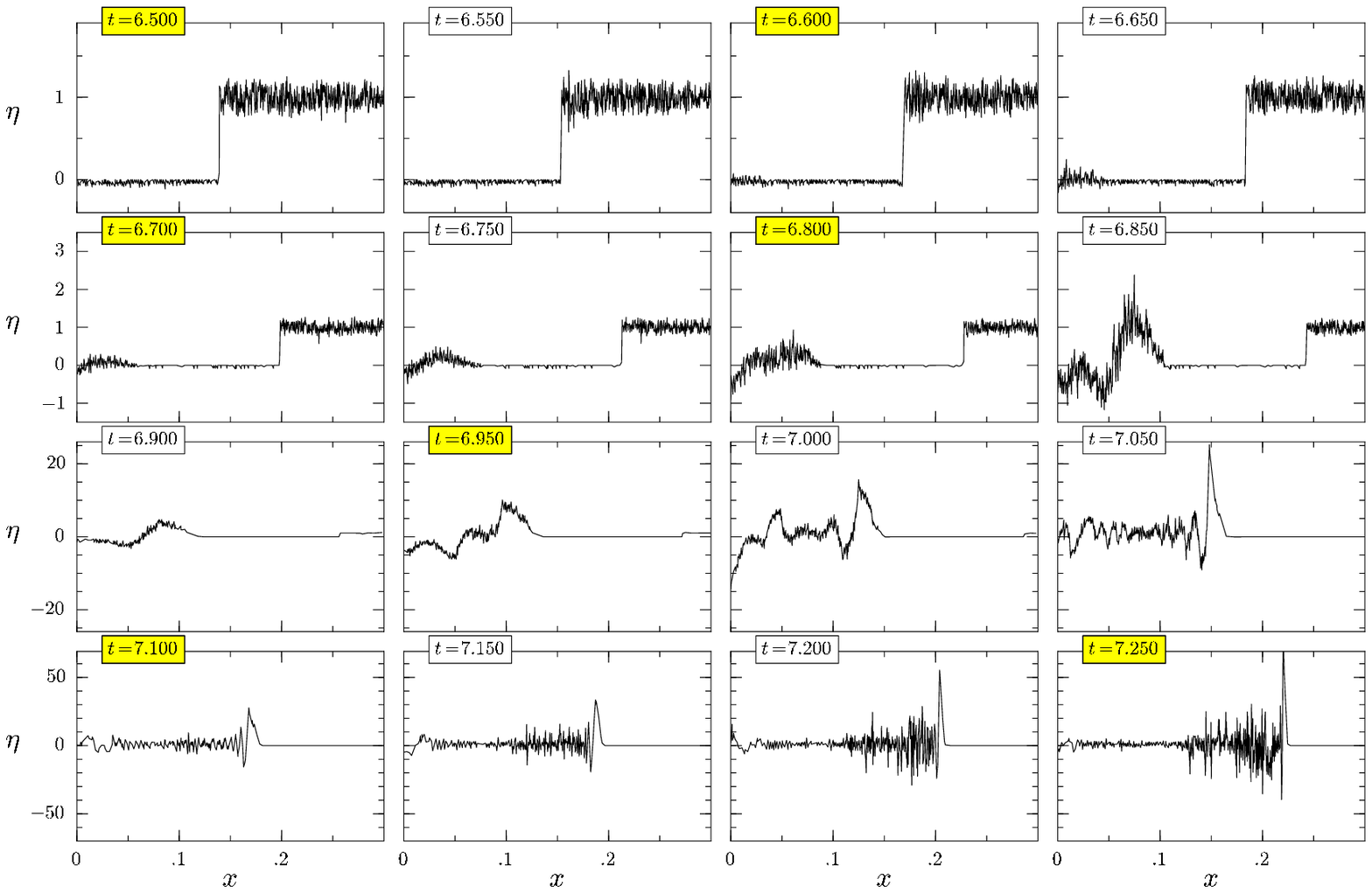}
\end{center}
\caption{Snapshots of charge density distribution in the calculation
  domain for cascade with $\jm=\GJ{j}$.  Charge density $\eta$  as a function
  of distance $x$ from the NS is plotted at equally separated
  moments of time;  $\eta$ is normalized to the Goldreich-Julian
  change density $\GJ{\eta}$.  The time $t$ shown in small 
  square boxes is normalized to the flyby time of the computation
  domain and is counted from the start of the simulation.  The presented
  cycle is taken from the middle of a long simulation.  
  \textbf{top}: The \emph{whole} cycle of cascade development.
  \textbf{bottom}: Snapshots for time interval marked by the gray
  area in the top panel;  these snapshots illustrate formation and
  propagation of plasma blob in more detail.
  \label{fig:tss_j1}}
\end{figure*}

\begin{figure*}
\begin{center}
  \includegraphics[width=\textwidth]{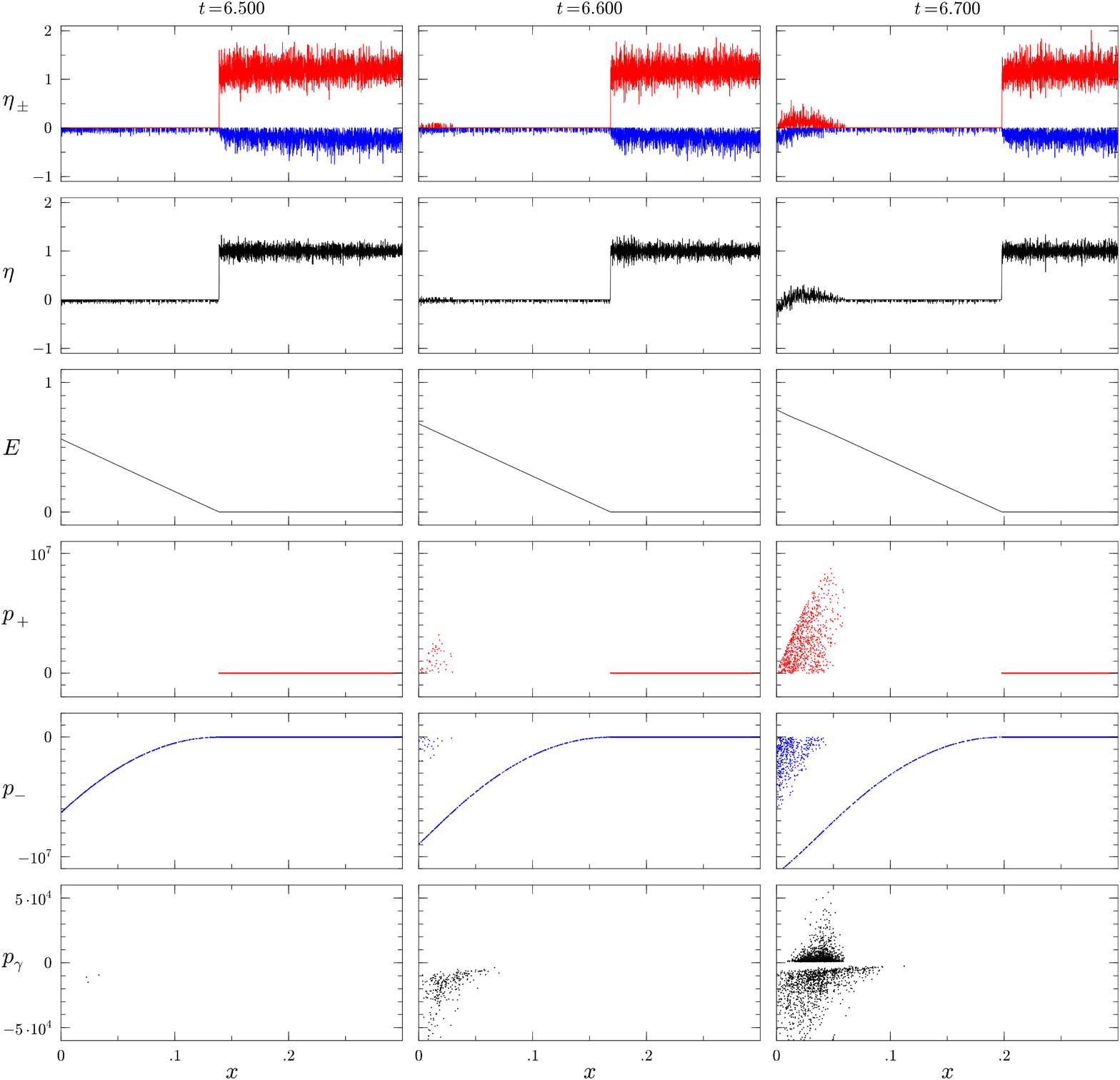}
\end{center}
\caption{Ignition of pair formation in cascade with $\jm=\GJ{j}$.
  Several physical quantities are shown as functions of the
  distance $x$ from the NS.  Plots in each column (for the same time
  $t$) are aligned -- they share the same values of $x$. 
  Snapshots are take at time moments of the first three marked
  snapshots in the bottom panel of Fig.~\ref{fig:tss_j1}.  
  The following quantities are plotted:
  \textbf{$\mathbf 1^{st}$ row:} $\eta_{\pm}$ -- 
  charge density of electrons (negative values, blue line) and
  positrons (positive values, red line);  
  $\eta_{\pm}$ is normalized to the Goldreich-Julian charge density $\GJ{\eta}$.
  \textbf{$\mathbf 2^{nd}$ row:} the total charge density $\eta$
  normalized to the Goldreich-Julian charge density $\GJ{\eta}$.
  \textbf{$\mathbf 3^{rd}$ row:} accelerating electric field
  $E$ normalized to the vacuum electric field $E_0$.
  \textbf{$\mathbf 4^{th}$ row:} phase space portrait of positrons
  (horizontal axis -- positron position $x$, vertical axis --
  positron momentum $p_{+}$ normalized to $m_ec$).
  \textbf{$\mathbf 5^{th}$ row:} phase space portrait of electrons
  (horizontal axis -- electron position $x$, vertical axis --
  electron momentum $p_{-}$ normalized to $m_ec$).
  \textbf{$\mathbf 6^{th}$ row:} phase space portrait of
  pair-producing photons 
  (horizontal axis -- photon position $x$, vertical axis --
  photon momentum $p_{\gamma}$ normalized to $m_ec$).
  \label{fig:ctss_j1_ignition}}
\end{figure*}

\begin{figure*}
  \begin{center}
    \includegraphics[width=\textwidth]{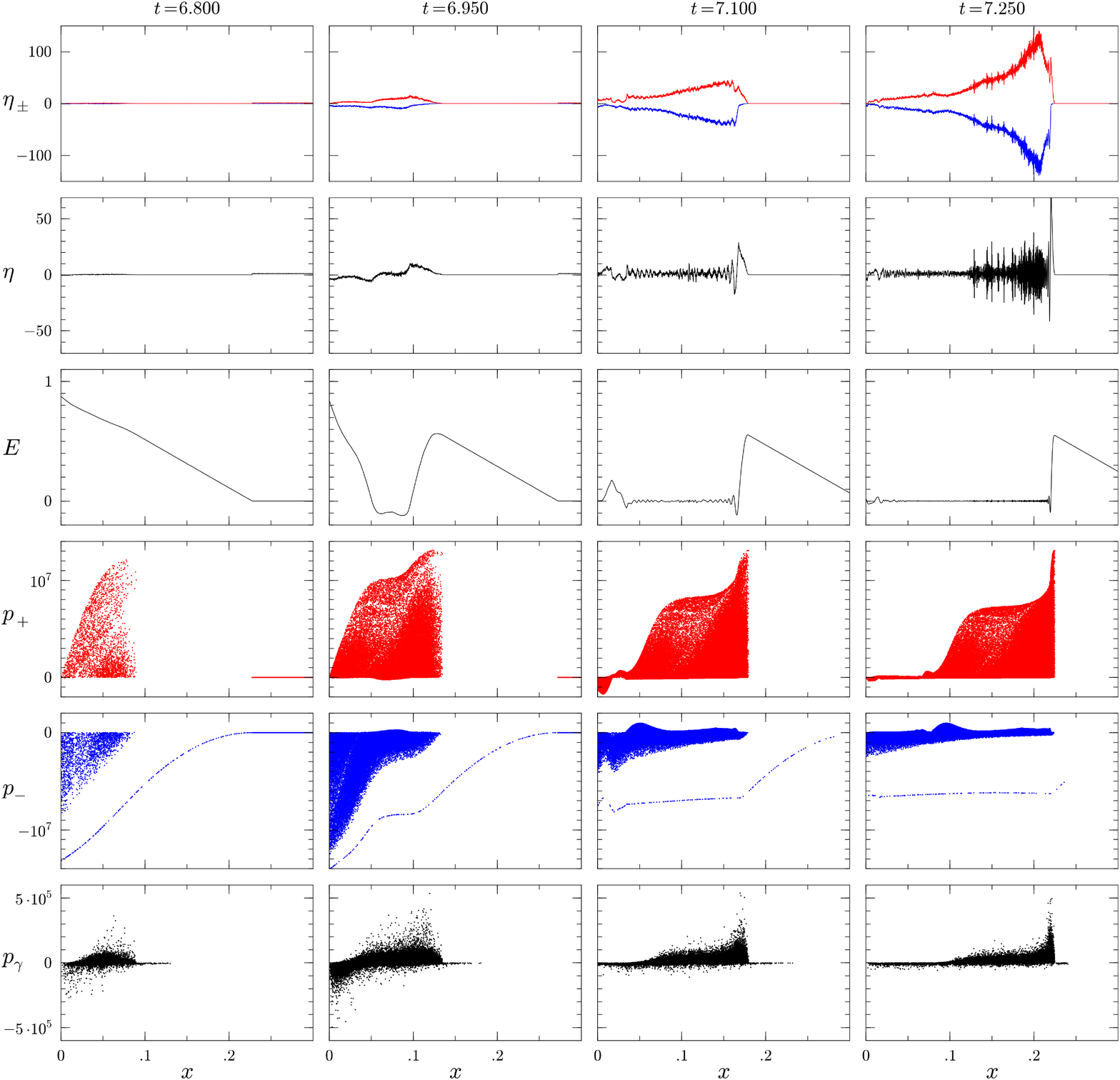}
  \end{center}
\caption{Screening of the electric field in cascade with
  $\jm=\GJ{j}$.   Snapshots are take at time moments of the last four marked
  snapshots in the bottom panel of Fig.~\ref{fig:tss_j1}.  The same
  quantities are plotted as in Fig.~\ref{fig:ctss_j1_ignition}. 
  \label{fig:ctss_j1}}
\end{figure*}

In this subsection I describe the pattern of plasma flow for a typical
cycle of pair formation in cascade with $\jm=\GJ{j}$.  Cascade
development is illustrated by a series of snapshots at several time
moments during a cycle of pair formation taken from a long simulation
where several such cycles were observed%
\footnote{In a previous short publication \citep{Timokhin_FERMI2009} I
  presented plots similar to Figs.~\ref{fig:tss_j1},~\ref{fig:ctss_j1}
  of this paper for a different cycle of the same simulation.
  Comparing these plots one can see that different bursts of pair
  formation are indeed very similar.}.
In Fig.~\ref{fig:tss_j1} I plot the change density at equally spaced
time interval during the discharge cycle.  In the upper panel of that
figure I present an overview of the entire cycle, in the lower panel I
plot snapshots of the change density distribution at smaller time
intervals for the most interesting part of the discharge -- formation
of a new plasma blob.  In
Figs.~\ref{fig:ctss_j1_ignition},~\ref{fig:ctss_j1} more detailed
information about physical conditions in the discharge zone is shown:
the number densities of electrons and positrons $\eta_\pm$, the
accelerating electric field $E$, phase portraits ($p-x$ diagrams) of
electrons, positrons and pair producing photons.  In the phase
portraits particles with positive values of 4-momentum $p$ are those
which move from the NS, particles with negative $p$ move toward the
NS.  The time $t$ in this figures is normalized to the flyby time of
the computational domain -- the time a relativistic particle needs to
cross the domain $L/c$.  The time is counted from the start of a
particular simulation, so its absolute value has no physical meaning
-- only time intervals between the shots have physical meaning.

Each pair creation cycle could be conveniently divided into three
phases: i) vacuum gap formation (timeshots for $t=6.033-6.633$ in
Fig.~\ref{fig:tss_j1}), ii) formation and propagation of a plasma blob
($t=6.633-7.833$) iii) relaxation ($t=7.833-8.833$).  Each burst of
pair formation generates dense electron-positron plasma which screens
the electric field.  Particles must leave the domain in order to
provide the required current density.  When plasma leaves the polar
cap a gap with almost no particles inside is formed; the vacuum
electric field in the gap is no longer screened (phase (i)).  The few
particles in the gap are accelerated and emit high energy
gamma-photons, and the process of plasma creation starts again.
Electron-positron plasma is produced non-uniformly, it forms a blob%
\footnote{Actually I am computing plasma sheets in 1D, but in 2D and
  3D these would be plasma "blobs", so I use the latter term
  throughout the paper.}
of relativistic plasma where large amplitude plasma oscillations are
excited (phase (ii)); the blob is visible in Fig.~\ref{fig:tss_j1} as
a packet of large amplitude charge density oscillations.  The blob
moves into the magnetosphere leaving a tail of moderately relativistic
plasma behind.  When the blob leaves the computational domain, the
remaining plasma still screens the vacuum electric field till the
plasma density drops below $\GJ{n}$ and pair formation starts again
(phase (iii)).  Below I describe these processes in more details.


A typical cycle starts with formation of a vacuum gap above the NS
surface (timeshots at $t=6.033-6.650$ in
Figs.~\ref{fig:tss_j1},~\ref{fig:ctss_j1_ignition}).  The gap forms
because plasma leaves the domain in order to provide the required
current density $\jm$.  The GJ charge density is positive, and when
the electric field is completely screened there are more positrons
than electrons.  The current density is positive, and positrons, on
average, must flow toward the magnetosphere (to the right).  As there
is no plasma inflow through the domain boundaries, the gap forms when
there are not enough charged particles to provide the required current
and change densities in the whole domain; above the gap the plasma
still sustain the required current and change densities.  Positrons
flow into the magnetosphere, so the gap forms at the NS surface.

Pair creation starts close to the NS and is ignited by gamma-rays
emitted by electrons flowing \emph{toward} the NS.  These primary
electrons have been created in the previous burst of pair formation,
they leak from the tail of the plasma blob formed in the previous
cycle and enters the gap from above.  These electrons are visible as a
thin line of particles with negative momenta $p_{-}$ in the electron
phase space portraits (the $5^{th}$ row in
Fig.~\ref{fig:ctss_j1_ignition}).  Pair production capable
gamma-photons emitted by these electrons have negative momenta and are
visible in the photon $p-x$ diagrams as scattered dots with negative
$p_\gamma$ at $t=6.5,6.6$.  These photons are eventually absorbed and
create electron-positron pairs, which are visible as scattered dots at
the left in the electron and positron $p-x$ diagrams at $t=6.6-6.7$.
These newly created electrons and positrons are accelerated by the
strong electric field of the gap (see timeshot at $t=6.6$); when they
have been accelerated up to Lorentz factors of the order of several
$10^6$ they start to emit pair production capable photons.  At
$t=6.700$ there are already gamma-photons with positive momenta, they
have been emitted by the secondary positrons accelerated in the strong
vacuum electric field.  At these early stages of pair discharge the
density of the newly created plasma is still very low (see plots for
$\eta_{\pm}$, the $1^{st}$ row in Fig.~\ref{fig:ctss_j1_ignition}),
and the electric field is not influenced by the injected plasma (the
$3^d$ row in Fig.~\ref{fig:ctss_j1_ignition}).  All electrons and
positrons are accelerated up to high Lorentz factors and start
emitting pair production capable gamma-photons very soon.


The secondary electrons and positrons are accelerated in opposite
directions, and plasma start being polarized (see distribution of the
charge density at $t=6.8,6.85,6.9$ in Fig.~\ref{fig:tss_j1}).  The
polarization of plasma creates an electric field opposite to the
vacuum field, and the effective accelerating electric field decreases.
When the particle number density become comparable to the
Goldreich-Julian density $\GJ{n}$ the accelerating electric field
starts being screened by plasma.  How the electric field is screened
is shown as a series of snapshots in Fig.~\ref{fig:ctss_j1}.  The
screening naturally starts in the place where plasma density is
maximal.  When more and more plasma is injected the region of screened
electric field broadens till it eventually extends up to the NS
surface.

The particles which produce the most of the pair creating photons are
the secondary positrons which have been accelerated at the time when
plasma density was small and electric field was strong.  Secondary
electrons have been accelerated up to very high energies too, but they
have been moving toward the NS -- they slammed into the NS surface and
do not contribute to pair creation at later times.  The high energy
positrons move into the magnetosphere emitting gamma-rays which turn
into electron-positron pairs.  These positrons practically co-move
with the gamma-rays.  Freshly created pairs are relativistic and have
momenta directed from the NS; most of them will remain relativistic
and will move into the magnetosphere.  So, the pair plasma forms a
blob with constantly increasing particle density (see the $1^{st}$ row
in Fig.~\ref{fig:ctss_j1}).

Photons cannot go ahead of relativistic particles at the front edge of
the blob, and pairs are injected only inside the blob.  As the plasma
blob moves into the magnetosphere so does the vacuum gap limited from
below by the front edge of the blob.  Ahead of the blob there is
practically vacuum with strong electric field, as there are only
electrons leaking from the tail of the previous blob and their number
density is very low.  The electric field inside the blob is screened
by the plasma.  At the front of the blob a sheath of positive charge
is formed which screens the vacuum electric field.  This sheath is
visible as a large spike in the charge density distribution at
timeshots $t=7.05-7.433$.  Inside the sheath the electric field goes
from the vacuum value to its very low value in the blob.  Pairs
injected in this sheath by conversion of gamma-photons emitted by the
particles which are already in the sheath are accelerated by that
electric field and start emitting gamma-rays too.  As more and more
pairs are ejected there the width of the sheath decreases.  However,
the number of particles in the sheath is small (see plot for
$\eta_{\pm}$ in Fig.~\ref{fig:ctss_j1} at $t=7.25$).  Pairs injected
in other parts of the blob are not accelerated because the electric
field there is screened.  Hence, subsequent pair formation is driven
mostly by particles accelerated at early stages of the discharge, when
the electric field was strong.  As these particles are emitting
gamma-rays and they are not accelerated anymore, their kinetic energy
decreases.  This can be seen in the positron $p-x$ diagrams in
Fig.~\ref{fig:ctss_j1}; the spike of high energy particles at the
blob's front is due to particle acceleration in the charge sheath.


I observe thermalization of freshly created electrons and positrons in
the simulations.  Low energy particles are present starting from very
early stages of blob formation.  Some of the low energy particles are
reversed back.  While the bulk of the plasma moves with relativistic
speed into the magnetosphere some particles are left behind forming a
``tail'' of the blob which has much lower particle number density than
the blob itself.  Although the fraction of particles left behind is
small, their number is enough to screen the vacuum electric field for
some time, preventing immediate formation of a new vacuum gap after
the blob detaches from the NS.

While the general structure of the flow is evident from the performed
simulations, some questions remain unanswered.  The most important
among them is about the time between discharges.  It depends on the
rate of plasma leakage from the blob -- the more particles leak out,
the later the next gap forms.  Due to continuous pair injection plasma
density in the blow increases enormously and at some time the
numerical scheme stops resolving the Debye length of the plasma, and
results start depending on the numerical resolution.  Because of this
the blob cannot be followed for time interval long enough (and
distances large enough) to get the repetition rate of the cascade.  In
the presented simulations the size of the simulation zone is set such
that the blob leaves calculation domain before the numerical scheme
fails to model it correctly.  When the blob leaves the calculation
domain particles are still leaking from it into the tail.  When the
blob is no longer in the computational domain, particle supply to the
tail is stopped and the time interval during which plasma density in
the domain drops to the GJ density -- and a new gap begins to form --
is substantially smaller in my simulations that it would be in
reality.  Duration of the relaxation phase in the simulations
(timeshots $t=7.833-8.833$) is strongly influenced by the numerical
setup.

However, I believe that the qualitative behavior of the plasma during
this phase is represented correctly, as there are no physical
processes in the blob tail except the particle supply from the blob
that the code fails to model when the blob is no longer in the
computational domain.  There is no particle acceleration in the tail
and pair creation is only due to electrons which leak from the
previous blob and are accelerated toward the NS in the traveling
vacuum gap.  The number of these electrons is negligible compared to
the number of pair-producing positrons in the blob.  The fraction of
particles leaking from the blob when it is still inside the domain is
also small, of the order of few per cents.  This is enough to screen
the vacuum electric field, but the energy carried by those particles
is negligible compared to the energy carried by particles in the blob
(see Sec.~\ref{sec:cascade-energetics}).  Hence, the only cascade
characteristic substantially influenced by the rate at which particles
leak from the blob is the repetition rate of pair creation bursts.
Already from this simulations it is clear that the time between
discharges is longer than the vacuum gap crossing time $\RS{h}/c$.
This introduces a new time scale into the Ruderman-Sutherland model.

Qualitatively the plasma flow after the discharge should be as
follows.  The tail consists of particles leaked from the blob; those
are mildly relativistic particles, some of them are trapped in plasma
oscillations, and, on average, the tail moves with a subrelativistic
velocity.  The vacuum gap is limited from above by the tail of the
previous blob: the plasma must move in order to support the current
density $\jm$; when the plasma density drops to values comparable to
the GJ density, the plasma cannot support the required current density
and the GJ charge density at the same time, a gap appears, and so the
tail ends in a vacuum gap.  Among the trapped particles in the tail
there are both electrons and positrons which move toward the NS.  The
electrons at the tail's end enter the gap and get accelerated toward
the NS -- they will be the primary particles in the next burst of pair
formation.  The positrons entering the gap are reversed and are sent
into the magnetosphere.  The upper boundary of the vacuum gap -- the
tail's end -- moves with a subrelativistic velocity, the front of the
blob is formed by ultra-relativistic positrons and moves
relativistically; therefore, the gap shrinks when the blob moves into
the magnetosphere.  The velocity of the blob's tail $v_{\rm{}tail}$ is
several per cents less than the speed of light,
$v_{\rm{}tail}\sim0.95-0.99\:c$.  Eventually the front of the new blob
catches the tail of the previous blob; the disappearance of the gap is
directly visible in the case of $\jm=0.5\GJ{j}$ (see
Sec.~\ref{sec:casc-with-different-j_0}).  Therefore, the
magnetosphere in the open field line zone will be filled with plasma
everywhere and starting at some distance from the polar cap there will
be no gaps in plasma spacial distribution.

\subsection{Superluminal plasma wave}
\label{sec:superl-plasma-wave}

\begin{figure*}
  \begin{center}
    \includegraphics[width=\textwidth]{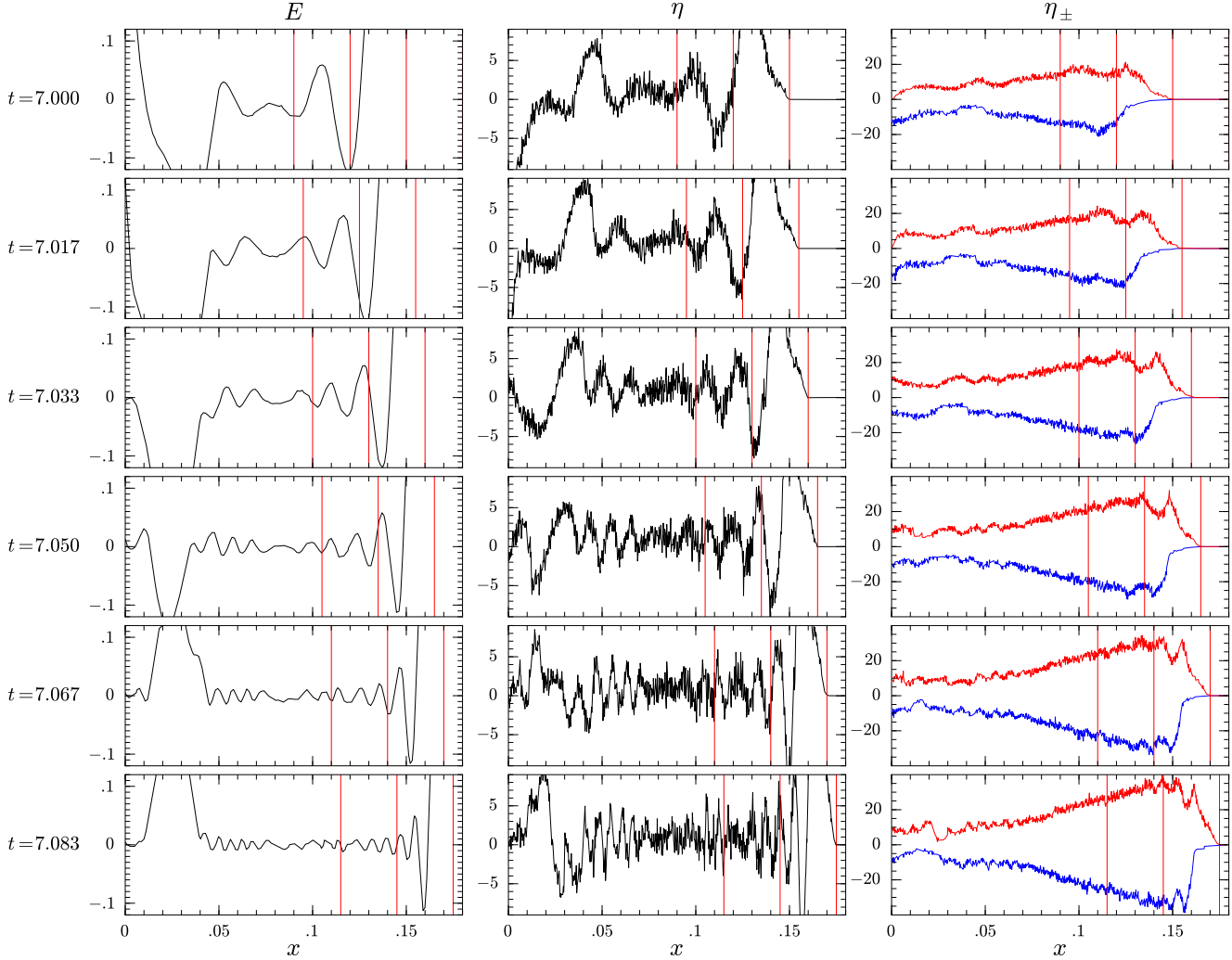}
  \end{center}
  \caption{Formation and propagation of superluminal electrostatic
    wave in the forming plasma blob for cascade with $\jm=\GJ{j}$.
    There are six snapshots for the electric field $E$, the total
    change density $\eta$ and the charge density of electrons
    (negative values, blue line) and positrons (positive values, red
    line) $\eta_{\pm}$. All quantities are plotted as functions of
    distance $x$ for the part of the calculation domain where the blob is 
    forming.  Snapshots are taken at equally separated time intervals.
    Plots in each column are aligned and share the same values of $x$.
    The same normalizations for physical quantities are used as in
    Fig.~\ref{fig:ctss_j1_ignition}.  The three thin red vertical
    lines in each plot mark fiducial points moving with the speed of
    light toward the magnetosphere.  The wave is superluminal, its
    maxima move faster that these lines.
    \label{fig:wave_propagation}}
\end{figure*}

When plasma starts being injected into a region with strong electric
field it is polarized and starts screening the electric field.  During
the process of screening of the vacuum electric field large-amplitude
oscillations are excited in the injected pair plasma; these
oscillations are visible in timeshots starting at $t=6.85$ and until
the blob leaves the domain.  Screening of the electric field starts in
the middle of the blob and spreads to its edges.  This spreading
occurs in the form of an electrostatic wave.  The propagation of the
wave can be seen in Fig.~\ref{fig:wave_propagation} where I plot
snapshots for the electric field, the charge density, and the particle
number density for the same spatial domain where the blob is being
formed for 6 moments of time; I plot also three vertical lines which
mark fiducial positions moving with the speed of light toward the
magnetosphere.  One can clearly see that the phase velocity of the
wave is greater that the speed of light; also note, that the wave
propagating toward the NS is superluminal too.  In the process of
electric field screening less and less charge separation would be
necessary to kill the electric field.  Apparently this is the reason
why the wavelength of plasma oscillations decreases.  At the time of
wave formation the particle number density is already very high, there
are more than $\sim100$ numerical particles per cell.  The Debye
length of the plasma -- calculated as
\begin{equation}
  \label{eq:lD_relativistic}
  \lambdaD \sim \frac{c}{\omega_p} = 
  c \left(\frac{4\pi{}e^2}{m_e}\int\frac{n(\gamma)}{\gamma^3}\,d\gamma\right)^{-1/2}\,,
\end{equation}
where $n(\gamma)$ is the number density of particles with the Lorentz
factor $\gamma$ -- is resolved; at the time the snapshots shown in
Fig.~\ref{fig:wave_propagation} are made $\lambdaD$ is several tens
times larger than the cell size.  Hence, these oscillations are not
numerical artifacts.  The electric field in the wave is too weak to
accelerate particles up to energies when they can emit pair production
capable photons.  Particle injection rate is set by very high energy
particles accelerated at earlier time and there is no back reaction of
the wave on the particle production rate.

Decreasing of the wavelength eventually ends when the wavelength
become equal to the cell size.  From that moments the code cannot
correctly follow propagation of the wave.  In timeshots for $t>7.25$
the wavelength is not resolved anymore.  Oscillations persists, but
because the wavelength is equal to the cell size, parameters of the
wave depend on the numerical resolution and so tell us little about
how the wave would propagate at that time in a real cascade.  However,
based on the information obtained at the early stages of wave
evolution (for $t<7.25$) -- when results of numerical modeling should
be reliable -- I would like to make the following remarks.  Being
superluminal this wave is not damped via Landau damping.  I may
speculate that this wave could stay superluminal, with its phase speed
approaching the speed of light from above, for a time long enough that
it can travel into the magnetosphere practically undamped.  Although
in my 1D simulations the wave is electrostatic, in reality it would be
electromagnetic.  It could escape the magnetosphere and be observed as
coherent pulsar emission, or/and it can excite another electromagnetic
wave(s) which escape the magnetosphere.

\subsection{Particle momentum redistribution and current adjustment}
\label{sec:part-moment-redistr}

\begin{figure*}
  \begin{center}
    \includegraphics[width=\textwidth]{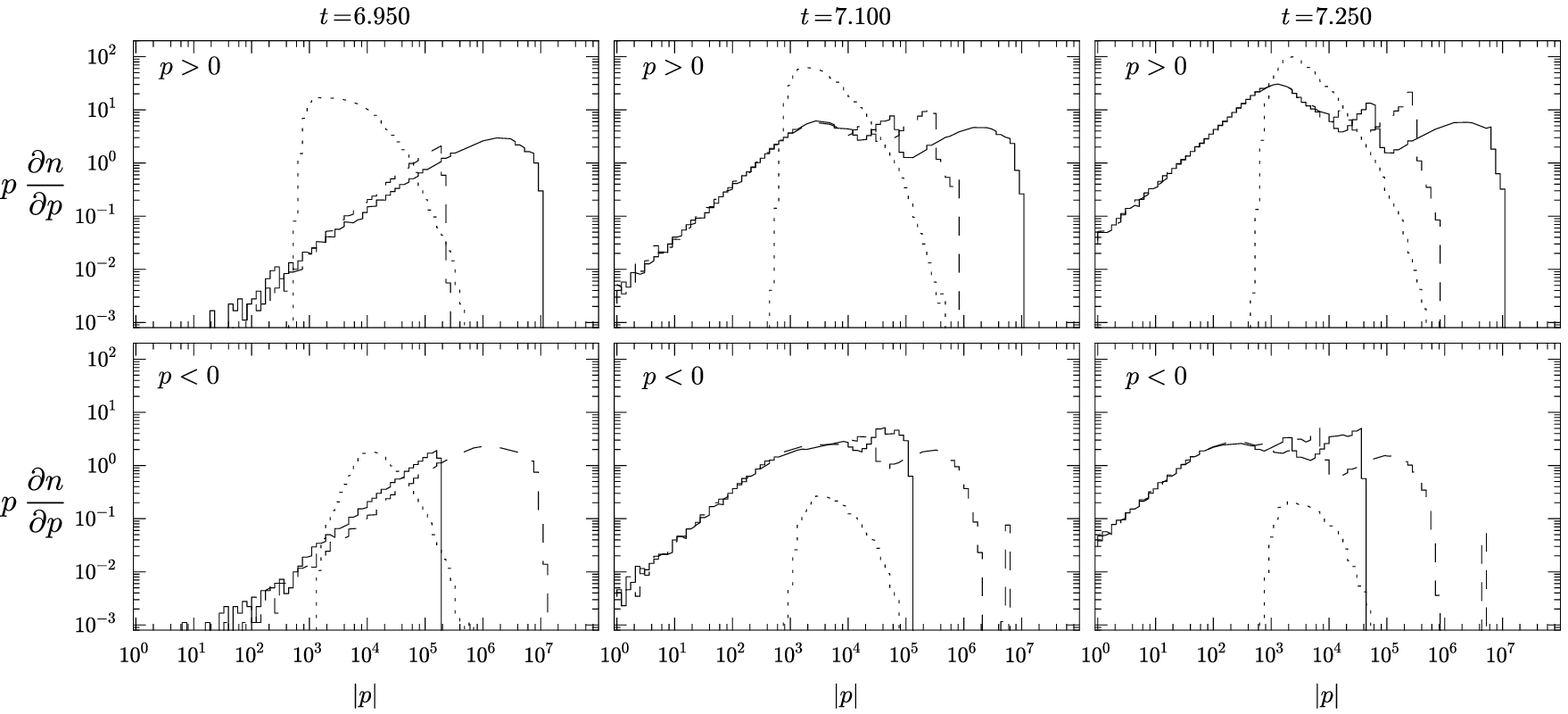}
  \end{center}
\caption{Particle momentum distribution for plasma in the blob at three
  moments of time for cascade with $\jm=\GJ{j}$. Positron
  distributions are shown by solid lines, electron distributions --
  by dashed lines, distribution of pair producing photons  -- by dotted
  lines. Plots in the top row show distributions for particles
  moving toward the magnetosphere ($p>0$), 
  plots in the bottom row -- distributions for particles moving toward
  the NS ($p<0$).  Each column corresponds to the same moment of time shown
  above the plots.  Plots in each columns are aligned and share the
  same values of $|p|$.  The following blob sizes were assumed:
  $x\in[0,0.135]$ for $t=6.95$, $x\in[0.05,0.18]$ for $t=7.1$ and
  $x\in[0.1,0.225]$ for $t=7.25$. 
  \label{fig:seds_j1}}
\end{figure*}

Starting from very early stages of blob formation the injected pairs
start being thermalized.  I use the term ``thermalization'' here
rather loosely, meaning that in the particle momentum distribution
there is a strong broad component which peaks at some momentum value;
it extends up to very low energies decreasing like a power law, and
decreases strongly after the peak.  Although I did not perform a
formal fitting procedure for particle momentum distribution%
\footnote{To get an accurate momentum distribution it is necessary to
  collect enough numerical particles. This leads to averaging over a
  macroscopic volume with different physical conditions, and results
  of such fitting would be ambiguous anyway},
the low energy tail of the ``thermal'' component follows the 1D
Maxwell-Juttner distribution $\partial{}n/\partial{}p\sim\mbox{const}$
for small $p$ quite good.  In Fig.~\ref{fig:seds_j1} I plot particle
momentum distribution $p\:(\partial{}n/\partial{}p)$ for three
different moments of time.  In the upper panel I plot the momentum
distribution of particles moving toward the magnetosphere, $p$ is
positive; in the lower panel -- the momentum distribution of particles
moving toward the NS, $p$ is negative.  These distributions are for
particles in the blob -- they are averages over $x\in[0,0.135]$ for
$t=6.95$, $x\in[0.05,0.18]$ for $t=7.1$ and $x\in[0.1,0.225]$ for
$t=7.25$ (cf. plots for the particle number density in
Fig.~\ref{fig:ctss_j1}).  There are low energy particles and there are
particles with both direction of motion in the blob.  Essentially the
pair plasma become four-component: (i) positrons moving into the
magnetosphere, (ii) positrons moving toward the NS, (ii) electrons
moving into the magnetosphere, (iv) electrons moving toward the NS.
Such four-component plasma easily adjusts locally to both requirements
of providing the GJ charge density and the current density $\jm$.

\begin{figure}
  \begin{center}
    \includegraphics[width=\columnwidth]{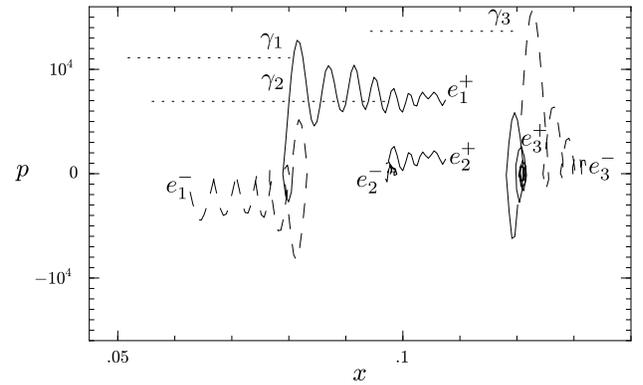}
  \end{center}
\caption{Trajectories of three pairs and their parent photons in the phase space
  (distance $x$ is along the horizontal axis, momentum $p$ -- along
  the vertical axis) for cascade with $\jm=\GJ{j}$.  Trajectories of
  pair creating photons are shown by 
  dotted lines and marked as $\gamma_{1,2,3}$.  Trajectories of
  positrons are shown by solid lines and marked as $e^+_{1,2,3}$,
  trajectories of electrons are shown by dashed lines and marked as
  $e^-_{1,2,3}$.  Marks are located at the ends of corresponding particle
  trajectories.  All trajectories end at $t=7.1$.
  \label{fig:trajectories_j1}}
\end{figure}

At initial phases of blob formation at least one of the processes
leading to plasma thermalization is trapping of particles in the
electric field of the plasma wave.  In Fig.~\ref{fig:trajectories_j1}
trajectories for three of such trapped pairs are shown.  In that
figure phase trajectories of pair-producing photons are plotted by
dotted lines, marked as $\gamma_{1,2,3}$.  When photon is absorbed an
electron and a positron are injected%
\footnote{Note that the initial kinetic energy of injected pairs is
  much less than the energy of the pair producing photons; this is
  easy to see although from eq.~(\ref{eq:p_injected}).  The rest of
  the energy goes into synchrotron radiation; that energy is carried
  by many photons with energies much less than that of the primary
  photon.}.
Electron trajectories are shown by dashed lines, positron trajectories
by the solid lines; trajectories' final points are marked as
$e^-_{1,2,3}$ and $e^+_{1,2,3}$ correspondingly.  Particle
trajectories end at $t=7.1$; at that and earlier time both the Debye
length and the wavelength of the wave are well resolved, they are tens
times larger than the cell size, and there are many particles per
cell; particle trajectories are well resolved too.  Hence, the
thermalization is not a numerical artifact.  Thermalization of freshly
injected pairs proceeds also at later stages of cascade development,
however, as the code does not resolve the plasma wave anymore, it is
not possible to disentangle the influence of the wave on the
thermalization process at this time.

From the current simulations it is not clear what is the fate of the
plasma wave in the blob.  If it exists for a long time, particle
thermalization in these oscillations could continue.  On the other
hand, deviation of the current density from $\jm$ results in
appearance of an electric field even if the charge density is equal to
the local GJ charge density.  The presence of that electric field in a
plasma with broad momentum distribution might result in some
instability which could facilitate pair thermalization -- at least I
see plasma thermalization during the relaxation phase when there is no
large amplitude plasma wave and plasma leaves the domain adjusting to
the required current density by redistribution of particle momenta.
This topic, however, needs additional investigation and will be
addressed in future publications.

To reverse the direction of motion of low energy particles a weak
electric field would be sufficient.  For charge density to be equal to
the local GJ charge density the number of positrons should be greater
that the number of electrons by $\GJ{n}$.  In order to provide current
density less than $\GJ{j}$ some positrons should move toward the NS.
If there is a population of low energy trapped positrons, a weak
fluctuating field can ensure that some trapped positrons \emph{on
  average} will be moving toward the NS.  A similar process can
provide a current density larger than $\GJ{j}$ and still keep the
change density equal to the GJ charge density.  The adjustment of the
current density can proceed in the cascade zone \emph{locally},
without inflow of charged particles from the magnetosphere (the latter
mechanism of current adjustment was discussed in
\citet{Lyubar92,Timokhin2006:MNRAS1}).

\begin{figure}
  \begin{center}
    \includegraphics[width=\columnwidth]{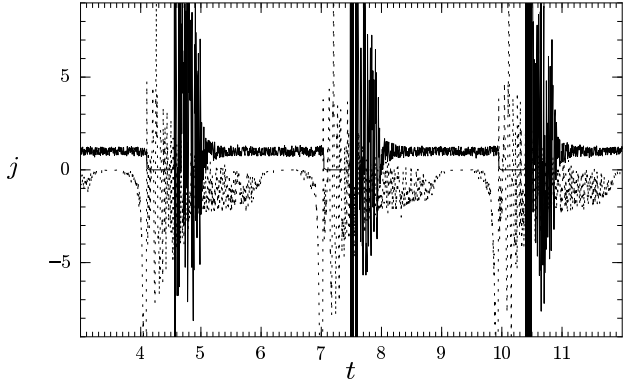}
  \end{center}
\caption{Currents through the domain boundaries for cascade with
  $\jm=\GJ{j}$ as functions of time for three consecutive bursts of
  pair formation.   The current flowing into the magnetosphere is shown by the solid
  line.  The current flowing \emph{into} the NS
  is shown by the dashed line (note that the direction for this current is
  opposite to the current direction assumed in the rest of the paper, so it should be
  on average negative).  Currents are normalized to the Goldreich-Julian current
  $\GJ{j}$. The currents are averaged over 10 time steps.
  \label{fig:j_flux_j1}}
\end{figure}

In Fig.~\ref{fig:j_flux_j1} I plot electric currents through the lower
domain boundary toward the NS surface (dashed line, this current
should be negative) and through the upper end of the calculation
domain (solid line, this current should be positive) as functions of
time.  The required current density $\jm$ is achieved on average, at
each moment of time the current density deviates from $\jm$.
Fluctuations are the strongest when particles from the burst of pair
formation hit the NS surface, and when the blob reaches the outer
boundary.  In all cases the relative deviation of the mean over the
cycle current density from $\jm$ is less than $\sim10^{-3}$.

\subsection{Cascade energetics}
\label{sec:cascade-energetics}

\begin{figure}
\begin{center}
  \includegraphics[width=\columnwidth]{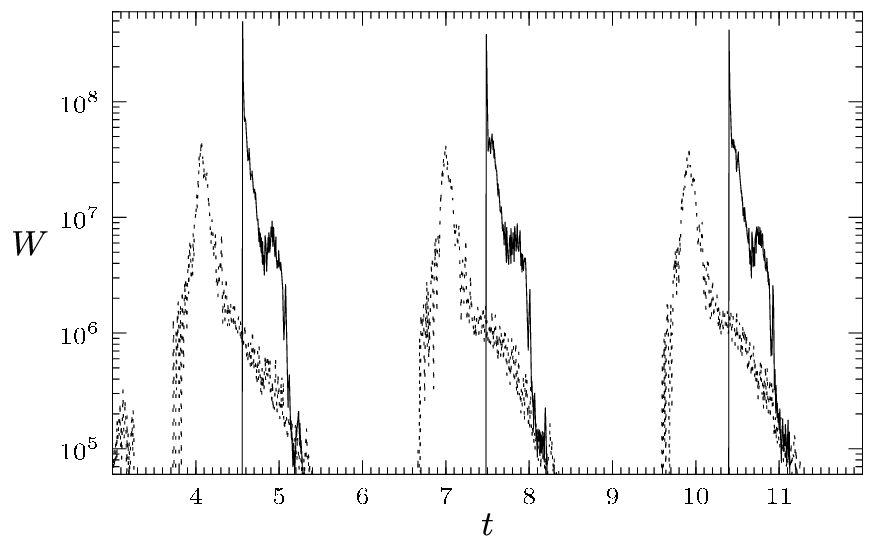}
\end{center}
\caption{Energy fluxes trough the domain boundaries for  cascade with
  $\jm=\GJ{j}$ as functions of time for three
  consecutive bursts of pair formation.  The flux toward the
  magnetosphere is shown by the solid line; the flux toward the
  NS is shown by the dashed line.  Fluxes are normalized to
  $m_ec^2\,\GJ{n}c$ and are averaged over 10 time steps.
  \label{fig:e_flux_j1}}
\end{figure}

The height of the gap is $\sim2$ times larger than the estimate given
by eq.~(\ref{eq:h_RS_num}).  The reason for this is that the upper
boundary of the vacuum gap -- the end of the blob tail -- is moving
into the magnetosphere while the electrons which ignite the cascade
are moving toward the NS.  When these electrons arrive at the point
where they emit first pair production capable photons, the gap's upper
boundary has moved some distance into the magnetosphere and the gap
size is larger.

The electric field in the gap linearly increases toward the NS and the
first secondary particles are injected into the region with very
strong electric field.  A substantial amount of particles needs to be
generated before the vacuum electric field is screened.  In the
meanwhile the vacuum gap is still growing and freshly injected
particles are accelerated in a very strong electric field.  In the
considered case a noticeable number of particles reaches the
radiation-reaction limited Lorentz factor $\sim1.4\times10^7$ (see
snapshots at $t=6.8,6.95$ in Fig.~\ref{fig:ctss_j1}).  This energy is
$\sim4$ times higher than $\RSarg{\gamma}{1}^{\max}$ given by
eq.~(\ref{eq:gamma_max_RS_num}).  When the electric field is screened
these particles start losing their energy quickly and then for the
most of the pair-producing positrons the Lorentz factor do not exceed
$\gamma_L\sim8\times10^6$ (see snapshots at $t=7.1,7.25$ in
Figs.~\ref{fig:ctss_j1}).  $\gamma_L$ is the Lorentz factor of a
relativistic electron/positron which loose substantial amount of its
kinetic energy due to curvature radiation while moving a distance
comparable to the length of the computation domain:
$\gamma_L/\delta{}t\sim{}W_{\rm{}rr}$; $\delta{}t=L/c$ and
$W_{\rm{rr}}$ is the radiation-reaction term in the equation of
motion~(\ref{eq:dp/dt}), it is given by eq.~(\ref{eq:W_rr}). In terms
of the problem's parameters 
\begin{equation}
  \label{eq:gamma_2}
  \gamma_L\sim{}5.6\times10^4\rho_6^{2/3}\left(\frac{L}{c}\right)^{-1/3}\sim{}8\times10^6\,,
\end{equation}
it is still $\sim2$ times larger than $\RSarg{\gamma}{1}^{\max}$ given
by eq.~(\ref{eq:gamma_max_RS_num}).  The particle energy distribution
at high energies is quite flat, see Fig.~\ref{fig:seds_j1}.  So, pair
producing particles are more energetic than it is expected from simple
estimates.  Because of these the total number of pairs generated by a
\emph{single} burst of pair formation should be larger than that
assumed in ``standard'' Ruderman-Sutherland model.  In order to get
the final pair multiplicity detailed full cascade simulations are
necessary, what will be subject of a separate research.  The total
number of high energy particles with $\gamma>5\times10^5$ in a blob is
$\sim{}0.7\:\GJ{n}\PC{r}$ per cm$^2$ of the blob perpendicular
cross-section.

In Fig.~\ref{fig:e_flux_j1} I show energy fluxes trough the lower and
upper boundaries of the calculation domain as functions of time.  The
energy flux hitting the NS surface is shown by the dashed line, the
energy flux going into the magnetosphere -- by the solid line.  The
fluxes are normalized to $m_ec^2\,\GJ{n}c$; they are computed by
summing kinetic energies of all particles leaving the domain at every
time step.  These functions have spikes when secondary particles
accelerated at the early stage of blob formation pass trough the
corresponding boundaries.  The energy carried by particles in the blob
tail is negligible (intervals between the spikes) and so the energy is
deposited only during bursts of pair formation.  The energy flux into
the magnetosphere is larger because most of the secondary electrons
slam into the NS surface before they achieve the maximum possible
energy while secondary positrons can gain the maximum energy as they
fly away from the surface.  The mean energy flux going toward the NS
averaged over the duration of the spike, e.g. over $t\in[6.6,8]$ in
Fig.~\ref{fig:e_flux_j1}, is $\sim4\times10^6m_ec^2\,\GJ{n}c$, the
flux going into the magnetosphere averaged over the same time is
$\sim8\times10^6m_ec^2\,\GJ{n}c$.  If the time between two successive
discharges is $(\PC{r}/c)f$ (i.e. the time between discharges is $f$
times larger that polar cap flyby time) the average energy flux going
into heating of the NS is
$\sim{}1.5\times10^{22}f^{-1}\mbox{erg s}^{-1}\mbox{cm}^2$,
this would result in the polar cap temperature
$\PC{T}\sim4\times10^6f^{-1/4}$~K,
if the heat conductivity is neglected.  The flux going into the
magnetosphere is
$\sim{}3\times10^{22}f^{-1}\mbox{erg s}^{-1}\mbox{cm}^2$.
Obviously, when the time between successive bursts of pair formation
is large, the overall heating of the NS surface is significantly
reduced.

\subsection{Cascades with $\jm=0.5, 1.5\protect\GJ{j}$}
\label{sec:casc-with-different-j_0}

\begin{figure*}
  \begin{center}
    \includegraphics[width=\textwidth]{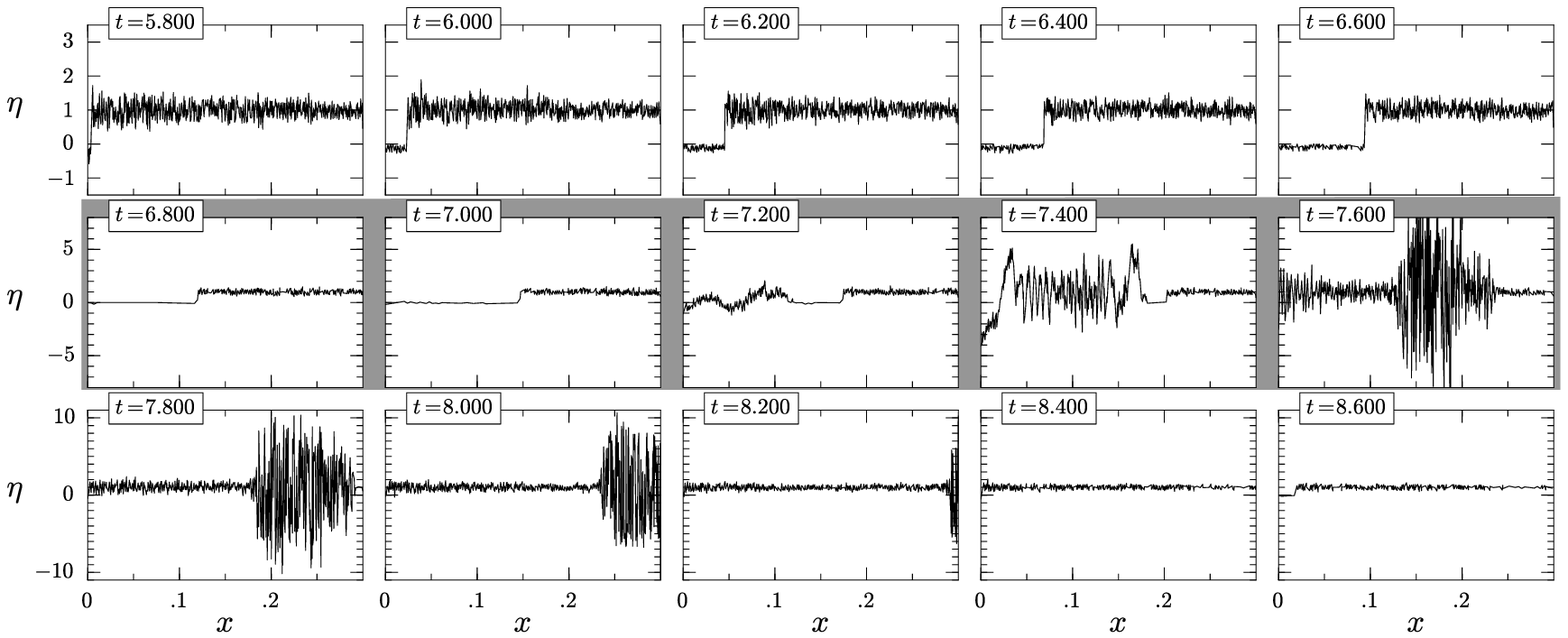}
    \includegraphics[width=\textwidth]{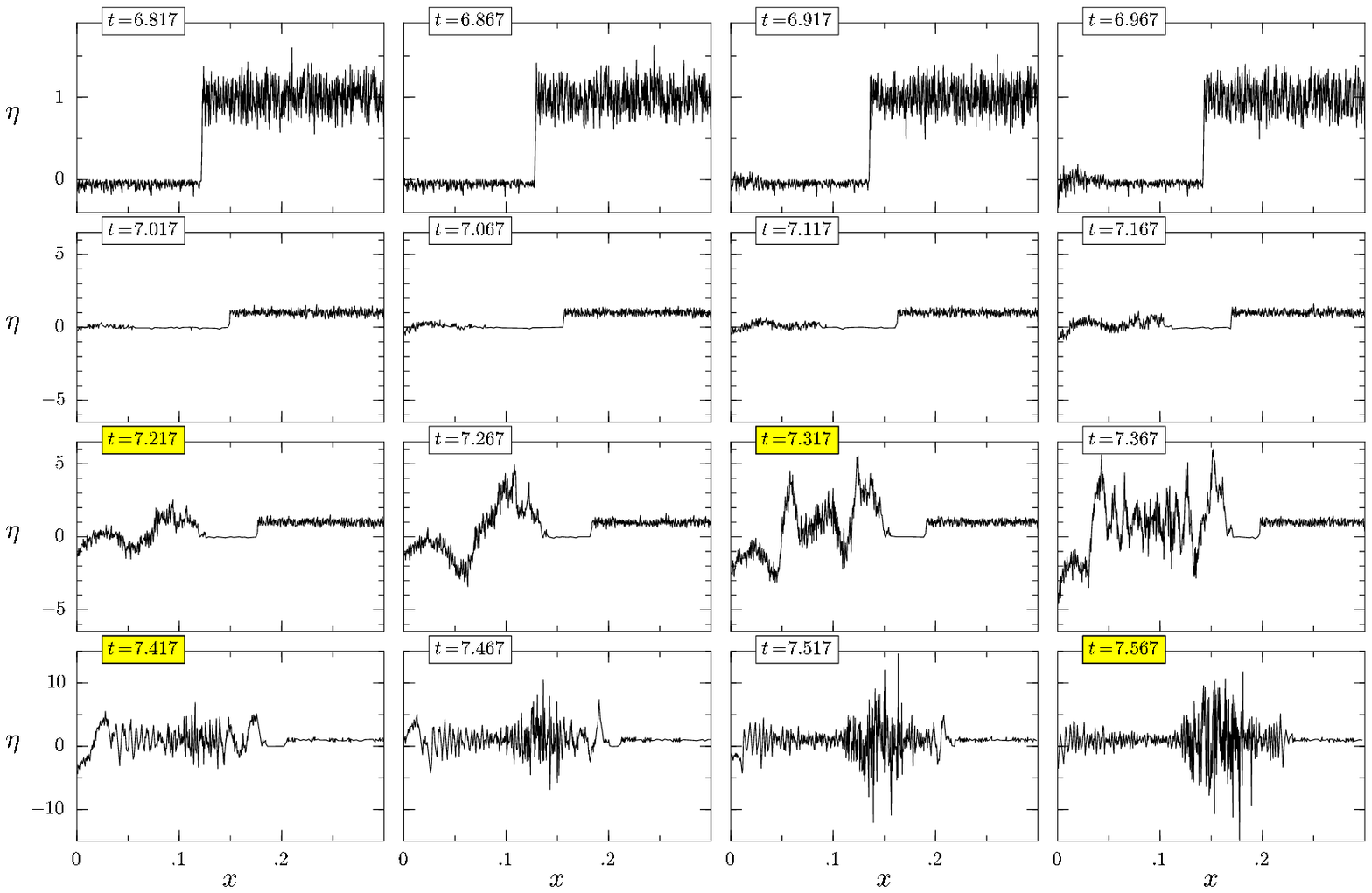}
  \end{center}
\caption{Snapshots of charge density distribution in the calculation
  domain for cascade with $\jm=0.5\GJ{j}$.  The same notations are
  used as in Fig.~\ref{fig:tss_j1}. 
  \label{fig:tss_j0.5}}
\end{figure*}

\begin{figure*}
  \begin{center}
    \includegraphics[width=\textwidth]{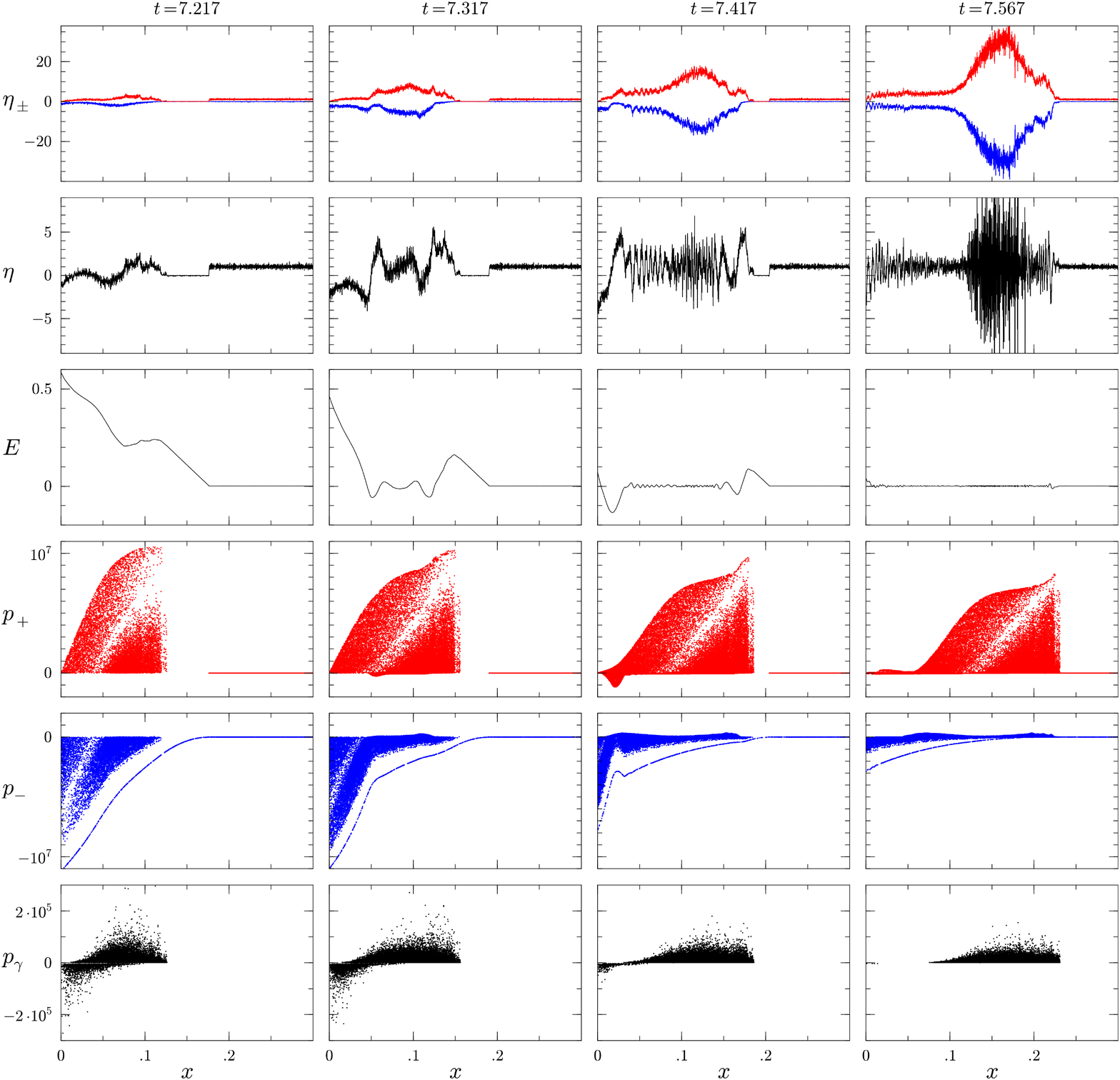}
  \end{center}
\caption{Screening of the electric field in cascade with
  $\jm=0.5\GJ{j}$.  Snapshots are take at the same time moments as the marked
  snapshots in the bottom panel of Fig.~\ref{fig:tss_j0.5}.  The same
  quantities are plotted as in Fig.~\ref{fig:ctss_j1_ignition}. 
  \label{fig:ctss_j0.5}}
\end{figure*}

\begin{figure*}
  \begin{center}
    \includegraphics[clip,width=\textwidth]{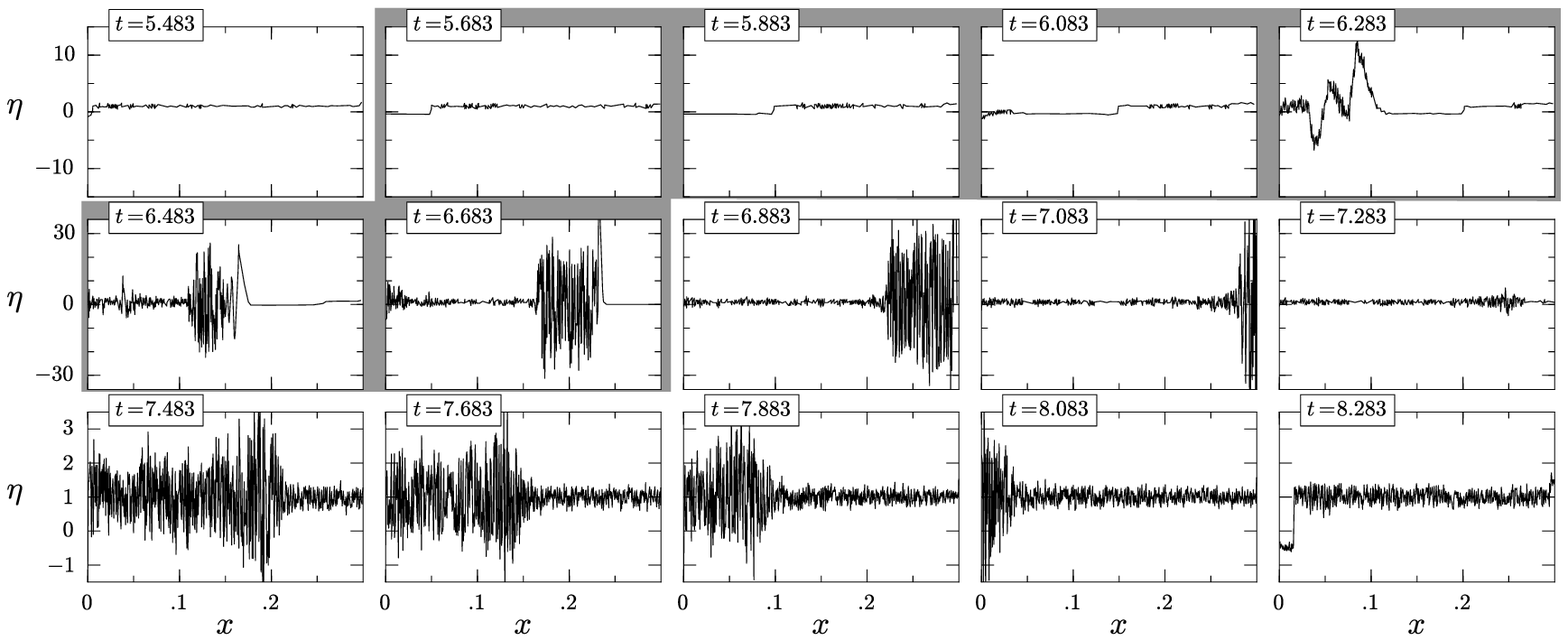}
    \includegraphics[clip,width=\textwidth]{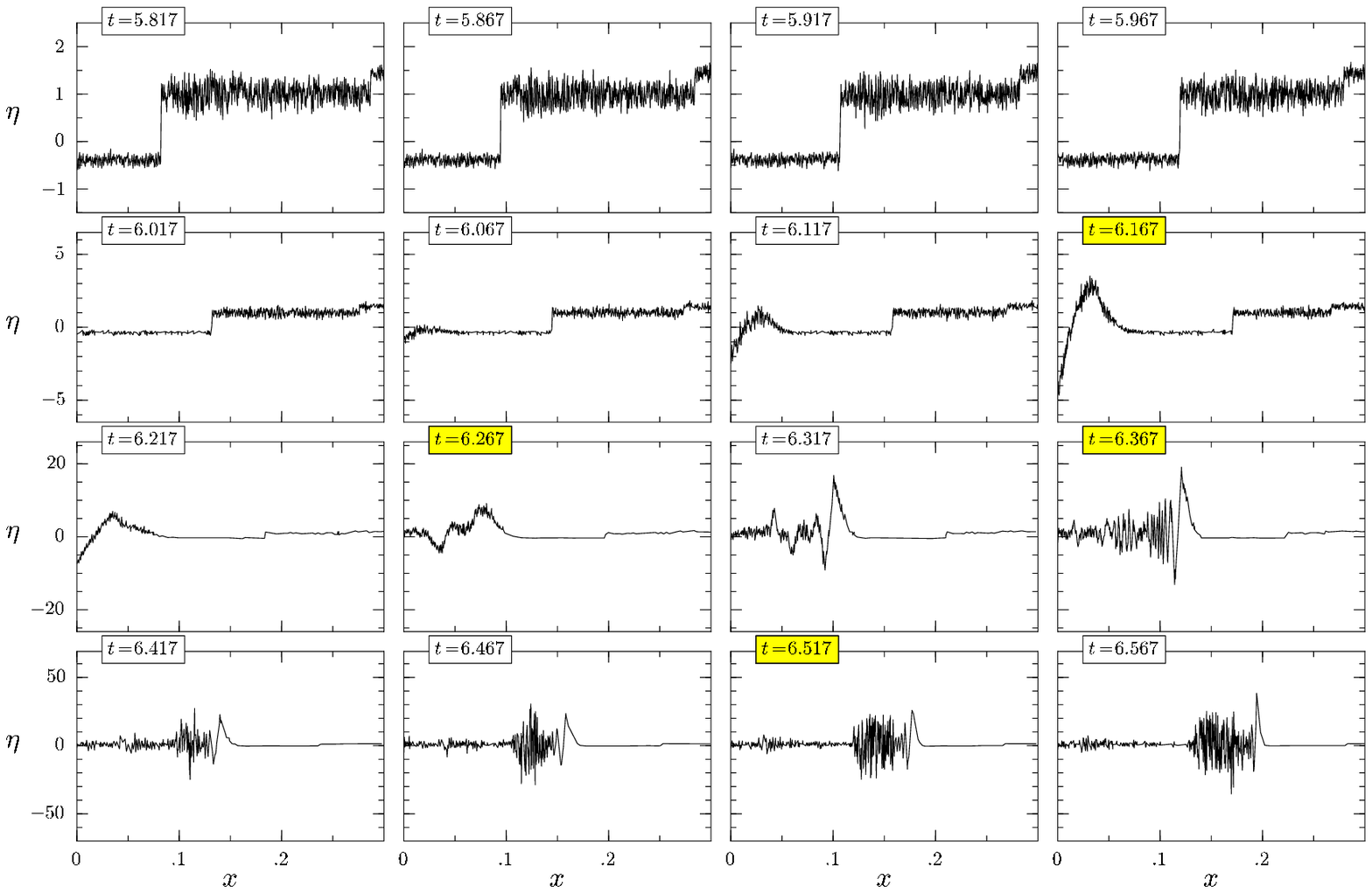}
  \end{center}
  \caption{Snapshots of charge density distribution in the calculation
    domain for cascade with $\jm=1.5\GJ{j}$.  The same notations are
    used as in Fig.~\ref{fig:tss_j1}.\label{fig:tss_j1.5}}
\end{figure*}

\begin{figure*}
  \begin{center}
    \includegraphics[clip,width=\textwidth]{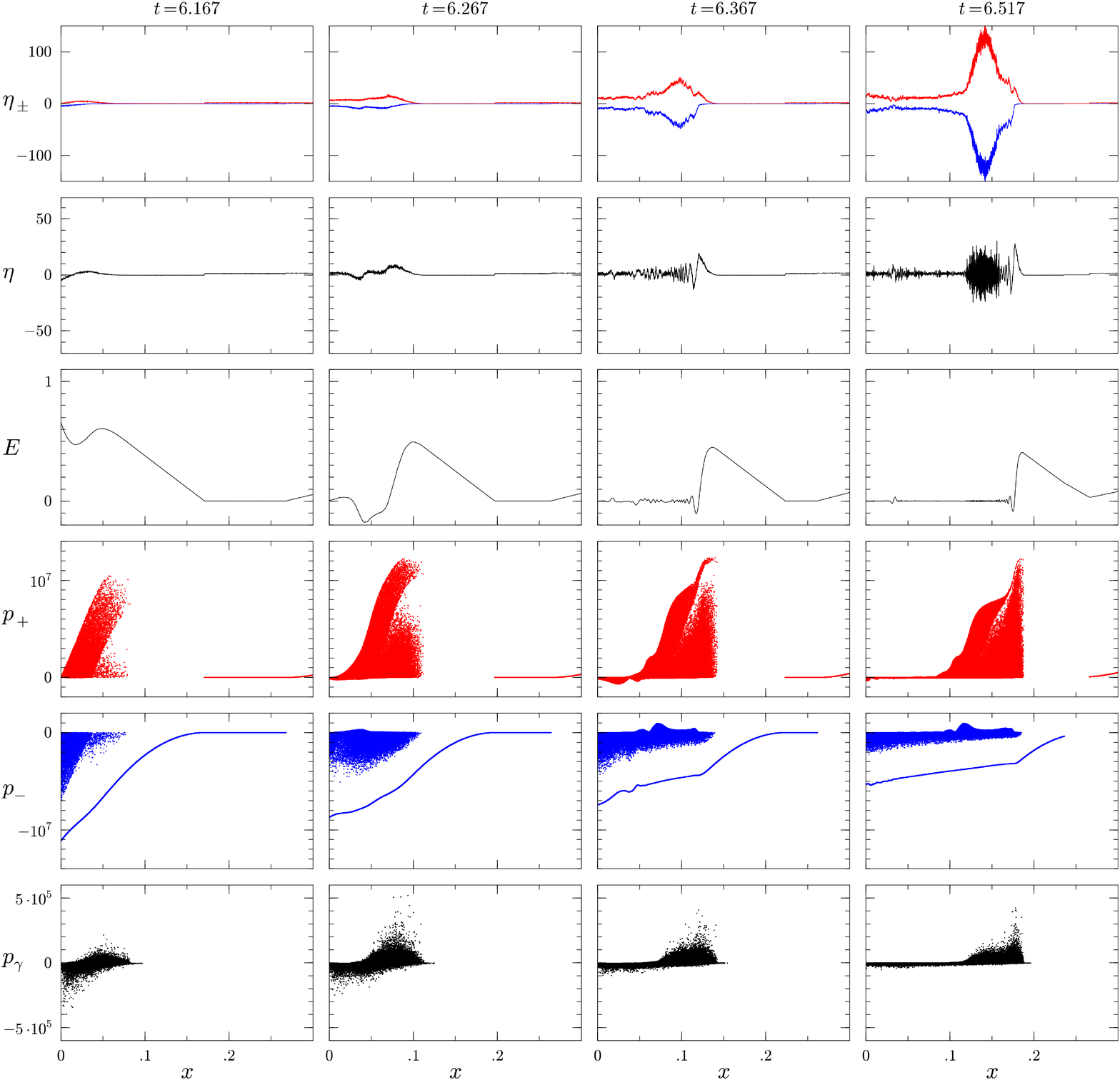}
  \end{center}
  \caption{Screening of the electric field in cascade with
    $\jm=1.5\GJ{j}$.  Snapshots are take at the same time moments as the marked
    snapshots in the bottom panel of Fig.~\ref{fig:tss_j1.5}.  The same
    quantities are plotted as in Fig.~\ref{fig:ctss_j1_ignition}. 
    \label{fig:ctss_j1.5}}
\end{figure*}

When the current density in the cascade zone is different from the GJ
current density plasma flow is qualitatively similar to the case
$\jm=\GJ{j}$ described above.  In Figs.~\ref{fig:tss_j0.5} and
\ref{fig:ctss_j0.5} I show snapshots of change density distribution
and detailed characteristics of cascade with $\jm=0.5\GJ{j}$; in
Figs.~\ref{fig:tss_j1.5},~\ref{fig:ctss_j1.5} -- the same plots for
cascade with $\jm=1.5\GJ{j}$.  The sizes of computation domains are
the same as in the case of $\jm=\GJ{j}$, so the time $t$ in these
plots, normalized to the flyby time, is measured in the same units.
The time intervals between individual plots in
Figs.~\ref{fig:tss_j0.5},~\ref{fig:tss_j1.5} are the same as in
Fig.~\ref{fig:tss_j1} discussed before.  Note, however, that snapshots
in Figs.~\ref{fig:ctss_j0.5},~\ref{fig:ctss_j1.5},~\ref{fig:ctss_j1}
are taken at different phases of the pair formation cycle.  In all
cases discharges repeat quasi-periodically and creation of pair plasma
goes through the same three phases: vacuum gap formation ($t=5.8-7$
for $\jm=0.5\GJ{j}$; $t=5.483-6.083$ for $\jm=1.5\GJ{j}$), formation
and propagation of the plasma blob ($t=7-8.2$ for $\jm=0.5\GJ{j}$;
$t=6.083-7.283$ for $\jm=1.5\GJ{j}$), and relaxation ($t=8.2-8.6$ for
$\jm=0.5\GJ{j}$; $t=7.283-8.283$ for $\jm=1.5\GJ{j}$).  The structure
of the plasma blob is similar -- there is a charged sheath screening
plasma in the blob from the large electric field in the vacuum gap,
and there are large amplitude plasma oscillations in the blob.  The
superluminal plasma waves are also present.

Cascade parameters for the case $\jm=0.5\GJ{j}$ differ from those for
the case $\jm=\GJ{j}$ as follows.  The size of the plasma blob is
larger. The velocity of the previous blob's tail is smaller,
$v_{\rm{}tail}\sim0.44c$, and the vacuum gap shrinks faster -- one can
see its disappearance in timeshots at $t=7.017-7.517$
(Figs.~\ref{fig:tss_j0.5},~\ref{fig:ctss_j0.5}).  When the gap closes,
the charge sheath at the blob front edge disappears and there is no
additional particle acceleration there -- in Fig.~\ref{fig:tss_j0.5}
at $t=7.467$ the sheath is still present, at $t=7.567$ it disappears.
The maximum particle energy is smaller -- particles do not reach the
radiation reaction limited energy -- but their maximum energy after
the electric field is screened is the same $\gamma_L$ given by
eq.~(\ref{eq:gamma_2}).  The plasma density in the blob is $\sim4$
times smaller.  The total number of high energy particles with
$\gamma>5\times10^5$ in the blob is $\sim{}0.22\:\GJ{n}\PC{r}$ per
cm$^2$ of the blob perpendicular cross-section, what is $\sim3.5$
times smaller than in the case $\jm=\GJ{j}$.  The energy flux toward
the NS is, as expected, smaller,
$5.8\times10^{21}f^{-1}\mbox{erg~s}^{-1}\mbox{cm}^2$, and the
estimated temperature of the polar cap is lower,
$\PC{T}\sim3.2\times10^6f^{-1/4}$~K.

Cascade parameters for the case $\jm=1.5\GJ{j}$ differ from those for
the case $\jm=\GJ{j}$ as follows.  The size of the plasma blob is
smaller.  The velocity of the previous blob's tail is
$v_{\rm{}tail}\sim0.82c$, it is smaller than that in the case of
$\jm=\GJ{j}$, but it is significantly larger than $v_{\rm{}tail}$ for
the case $\jm=0.5\GJ{j}$.  Although the gap shrinks faster it still
leaves the domain.  First generation secondary particles reach the
radiation reaction limited energy and then slow down to the Lorentz
factors $\la\gamma_L$.  The plasma density in the blob is slightly
higher.  The total number of high energy particles with
$\gamma>5\times10^5$ in the blob is $\sim{}0.6\:\GJ{n}\PC{r}$ per
cm$^2$ of the blob perpendicular cross-section, what is slightly less
than that value for $\jm=\GJ{j}$ cascade.  The energy flux toward the
NS is slightly lower,
$\sim1.3\times10^{22}f^{-1}\mbox{erg~s}^{-1}\mbox{cm}^2$, and the
estimated temperature of the polar cap
$\PC{T}\sim3.9\times10^6f^{-1/4}$~K.

These differences are ultimately related to the speed at which the
tail of the previous blob leaves the domain and how many electrons are
leaking from it.  The cascade energetics and the ultimate pair
multiplicity depends on the number of the first generation secondary
positrons, their maximum energy, and for how long time these particles
are sustained at this energy.  The first generation secondary
particles are accelerated up to very high energies -- they can be
accelerated up to the radiation-reaction limited energy -- and are
sustained at this energy until enough plasma is produced to screen the
electric field in the blob.  The rate of the first generation
secondary positrons production depends on how many electrons leak from
the tail of the previous blob, and the maximum energy of these
positrons depends on the electric field where they are injected -- the
faster the gap grow the larger the electric field.

In the case $\jm=\GJ{j}$ redistribution of particle momenta is not
necessary for adjustment of the current density -- bulk motion of the
tail toward the magnetosphere would provide the required current
density because the charge density is already $\GJ{\eta}$; so in this
case the average speed of the tail is the largest.  The gap grows
fast, and the first generation of secondary particles is created in a
region with very strong electric field.  When the current density
$\jm$ differs from the GJ current density, redistribution of particle
momenta is required to sustain $\jm$ by keeping at the same time the
charge density equal to the GJ charge density.  For $\jm>\GJ{j}$
electrons must be sent back to increase the current, for $\jm<\GJ{j}$
some positrons must be reversed to decrease the current.  The plasma
as whole moves into the magnetosphere; low energy particles are
trapped in small amplitude plasma oscillations and are dragged with
the bulk of the plasma.  Presence of a weak electric field would be
sufficient to ensure that the required number of particles on average
moves toward the NS.  Such particle reversal results in slower motion
of the tail.  When the gap upper boundary moves slower, the first
generation of secondary particles is injected in a weaker electric
field, and the overall energetics of cascade is lower.  On the other
hand, the larger the current density the larger the number of
electrons leaking from the tail; these electrons are the primary
particles igniting the cascade.  To screen the electric field at least
GJ number density of particles is required.  When the flux of the
primary electrons is higher, the number of the first generation
secondary pairs grows faster, and polarization of the plasma which can
screen the vacuum electric field is achieved at smaller spacial
separation; this results in a smaller size of the plasma blob.  So,
cascades with the current density equal to the GJ current density are
most energetic and should produce densest plasma.  However, as it
follows from the simulations, for $\jm>\GJ{j}$ cascade properties seem
to be less sensitive to the value of $\jm$ than properties of cascades
with $\jm<\GJ{j}$.  Hence, cascades with $\jm>\GJ{j}$ should have
energetics and final multiplicities lower but comparable to those of
cascades with $\jm=\GJ{j}$, while energetics and final multiplicities
of cascades with $\jm<\GJ{j}$ should be significantly lower.

\begin{figure}
  \begin{center}
    \includegraphics[width=\columnwidth]{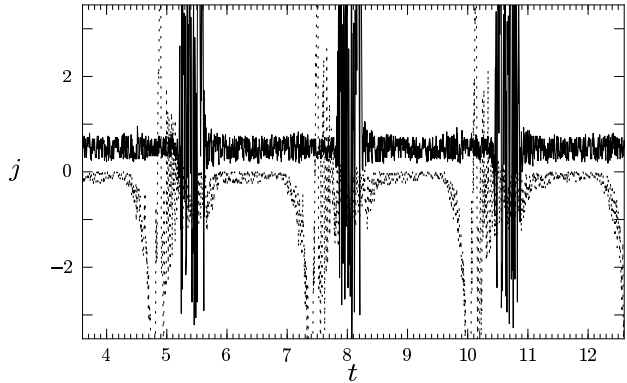}
  \end{center}
  \caption{Currents through the domain boundaries for cascade with
    $\jm=0.5\GJ{j}$ as functions of time for three consecutive bursts of
    pair formation. The currents are averaged over 10 time steps. 
    The same notations are used as in Fig.~\ref{fig:j_flux_j1}
    \label{fig:j_flux_j0.5}}
\end{figure}

\begin{figure}
  \begin{center}
    \includegraphics[width=\columnwidth]{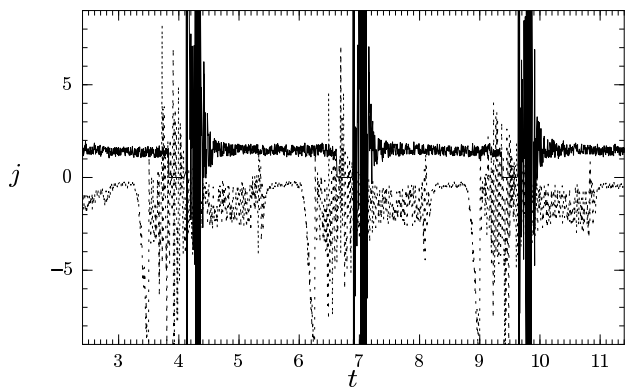}
  \end{center}
  \caption{Currents through the domain boundaries for cascade with
    $\jm=1.5\GJ{j}$ as functions of time for three consecutive bursts of
    pair formation. The currents are averaged over 10 time steps. The
    same notations are used as in Fig.~\ref{fig:j_flux_j1} 
    \label{fig:j_flux_j1.5}}
\end{figure}

In Figs.~\ref{fig:j_flux_j0.5} and \ref{fig:j_flux_j1.5} I plot
electric currents through the lower and upper domain boundaries for
cascades with $\jm=0.5\GJ{j}$ and $\jm=1.5\GJ{j}$ correspondingly.
Except for the value of the mean current density these currents behave
in a similar way as the currents for a cascade with $\jm=\GJ{j}$; the
relative deviation of the mean over the cycle current density from
$\jm$ is also less than $\sim{10^{-3}}$.

Regarding the repetition rate of pair formation bursts in cascades
with different current densities I can make only some qualitative
remarks.  For cascades with smaller $\jm$'s pair multiplicity is
smaller, what must result in a less dense plasma tail; there will be
less particles to wipe out of before the next vacuum gap can develop.
On the other hand, if the current density is smaller, particles are
wiped out slower because smaller current density requires less
particles to sustain it.  For cascades with higher current densities
pair multiplicity should be higher, but a larger particle flux is
required.  So, it seems that dependence of the time between the
discharges on the current density should be moderate; it is also
possible that this dependence is non-monotonic.

\subsection{Summary of cascade properties}
\label{sec:summ-casc-prop}

The case described in details in previous sections is a good example
of a typical Ruderman-Sutherland cascade.  Insights gained from
analyzing that case helped to draw a general physical picture.  I
performed simulation for different pulsar parameters ($P$, $B_0$,
$\rho$), and the results are in complete agreement with the general
picture outlined below.

Cascades show limit cycle behavior for all physical parameters
allowing pair creation.  Pair formation is quite regular -- in each
discharge a blob of pair plasma is formed, the blob propagates into
the magnetosphere leaving behind a tail of low energy particles.  When
particle number density in the tail becomes comparable to $\GJ{n}$ a
vacuum gap appears, and then a new blob of pair plasma forms.  There
are two characteristic timescales: i) time scale associated with the
size of the plasma blob $\tau_1=L_{\rm{}blob}/c$, ii) time between two
successive discharges $\tau_2$.  The first timescale is of the order
of $\tau_1\sim{}h_{RS}/c$ for all $\jm$, and it should be (much)
smaller that the second timescale, $\tau_1\ll\tau_2$.  The simulations
are inconclusive about the real dependence of $\tau_2$ on the current
density $\jm$, but it seems that this dependence might be weaker that
linear.

When a new bursts of pair creation occurs, the vacuum gap detaches
from the NS surface and propagates into the magnetosphere.  The upper
boundary of the gap moves subrelativistically, while the front of the
new blob moves ultra-relativistically; eventually the gap closes.  For
sub-GJ current densities the tail of the previous blob moves slower
and the gap disappears faster.  For cascades with super-GJ current
densities the gap should disappears faster for larger values of $\jm$,
but dependence of the tail velocity $v_{\rm{}tail}$ on $\jm$ is
significantly weaker than that in cascades with sub-GJ current
densities.  The gap should persists for longest time in cascades with
$\jm=\GJ{j}$ .

For pulsars with large potential drop across the gap the first
generation of pairs in cascades with $\jm\ga0.5\GJ{j}$ reaches the
radiation reaction limited energy.  When electric field is screened,
these particles propagate loosing their energy by emitting curvature
photons.  At first these losses are large, as radiation efficiency
depends on particle energy as $\propto\gamma^4$, but then particles
loose energy more slowly.  There is also a small amount of particles
in the charge sheath at the front of the plasma blob which feel the
non-screened electric field and are continuously accelerated.  For
current densities $\jm\la0.5\GJ{j}$ particles do not reach radiation
reaction limited energies.  However, in any case the kinetic energy of
the first-generation particles is larger than it follows from
estimates of \citet{Ruderman/Sutherland75} and so the final pair
multiplicity and energetics of a \emph{single} bust of pair formation
is higher.  The vast majority of energy is carried by the first
generation pairs and so the heating of the NS polar cap by the cascade
occurs in bursts, when first generation pair electrons hit the
surface; heating during the 'relaxation' phase is negligible.  If the
time between successive discharges is large, the heating will be much
lower than it is usually assumed in the Ruderman-Sutherland model.

During the discharge a superluminal electrostatic wave is formed.  As
its phase velocity is larger than the speed of light it is not damped
via Landau damping.  From the performed simulations the ultimate fate
of this wave is not clear because after some time the code stops
resoling its wavelength.

\section{Discussion}
\label{sec:discussion}

I performed for the first time self-consistent time-dependent modeling
of electromagnetically driven cascades which included all essential
physical processes.  I considered the simplest possible case -- the
Ruderman-Sutherland cascade, when all particles in the discharge zone
are produced by the cascade itself.  Cascade behavior is quite regular
-- spatially localized blobs of pair plasma are formed during
regularly repeating discharges, each blob propagates into the
magnetosphere leaving a tail of mildly relativistic plasma behind.
Energetics of individual discharges is higher than that predicted by
the Ruderman-Sutherland model.

The model is one-dimensional and includes the minimum set of processes
involved; but just because a ``bare-bone'' system was considered it
was possible to get insights about general physics of
electromagnetically driven cascades.  I was interested in dynamics of
electromagnetic discharges, i.e. how particle are accelerated when
pair production takes place.  I did not follow development of the full
cascade initiated by energetic particles moving in the magnetosphere
above the polar cap where there is no accelerating electric field.
The latter problem has been studied before by many authors, and
qualitative properties of cascades initiated by a relativistic
particle are relatively well known
\citep[e.g.][]{Daugherty/Harding82,Zhang2000,Hibschman/Arons:pair_production::2001,Medin2010}.
In future works I plan to address this problem using particle energy
distributions obtained from self-consistent discharge simulations.
Although quantitatively results of such full cascade simulations would
be different from that described in the above mentioned papers,
qualitatively they should be similar.

In my 1D simulations individual discharges are very similar and
electrostatic oscillations are clearly visible and coherent.  Usually
1D and 2D models show more coherent behavior than full 3D simulations
because of the inforced symmetry.  Hence, the coherent behavior
present in this 1D simulations could be somewhat washed out in a more
realistic 3D model.  The usual picture used in models of polar cap
discharges involves several separate discharge zones across the polar
cap -- ``sparks'' in terms of the Ruderman-Sutherland model.  Whether
interaction between sparks via induced electric field is strong enough
for them to be coupled is a priori not clear.  However, particle
motion in the superstrong magnetic field in the pulsar polar cap is
one-dimensional, the curvature of magnetic field lines is small, and
photon trajectories only slightly deviates from particles
trajectories; that suggests that most of the cascade properties
deduced from current simulations can be preserved in a full 3D model
for individual sparks, although only direct 3D simulations can prove
this.

I considered the case when particles cannot be extracted from the
surface of the NS, but the results may be applicable to a broader
class of problems.  If cascades under different physical conditions
work as series of discharges, their behavior should be similar to that
described here.  Namely, the blob-tail structure can be preserved,
pair plasma thermalization will take place, transient electrostatic
wave will be excited.  In particular, I suspect that polar cap
cascades with space charge limited flow -- in a non-stationary version
of the model suggested by \citet{Arons1979} and
\citet{Muslimov/Tsygan92} -- might have some similarities with the
case described here.

One of the main motivations to start this project was to study how
cascade zone provides the current density required by the
magnetosphere.  It has been speculated that non-stationary cascade can
be sustained at any average current density flowing through the
discharge zone \citep[e.g.][]{Levinson05,Timokhin2006:MNRAS1}.  My
simulations has shown explicitly that this is indeed the case.  The
current density at any given point fluctuates strongly, but on average
it is equal to the mean current density $\jm$ with very high accuracy.
Current adjustment works well in the relaxation phase too, when both
the charge and the current densities are close to the required values
even when they are averaged over timescales much smaller than the
flyby time of the calculation domain.  It works because in the tail
there are low energy particles trapped in small amplitude plasma
oscillations, and the weak fluctuating electric field of plasma
oscillations is able to reverse particles of \emph{both signs} (at
different oscillation phases).  Particles which flow backwards do it
in time averaged sense, they spend more time in backward motion; there
is no separate population of particles flowing only backwards all the
time.  Such way of current adjustment is possible because of effective
plasma thermalization which provides low energy particles.

The current density $\jm$ can have the sign opposite to the sign of
the GJ charge density, $\jm<0$ \citep{Timokhin2006:MNRAS1}.  I ran
simulations with $\jm<0$ too; everything works exactly in the same way
as in the case with $\jm>0$, except that the gap forms at the upper
domain boundary, at the magnetosphere' side.  This 1D problem is
symmetric in regard to the sign of the current density: if $\jm<0$,
the pair plasma -- which has positive net charge in order to sustain
the GJ charge density -- moves toward the NS and generates negative
$\jm$.  However, at the magnetosphere' side there is no solid surface
which can prevent charged particles from escaping; if a gap forms
there, some charged particles can be sucked from the magnetosphere.
In principle, it might result in presence of particles of both signs
in the gap and, therefore, in cascade ignition at both ends of the
gap; whether this is the case or not can not be decided based solely
on qualitative arguments and requires accurate quantitative modeling.
The gap will form in the tail of the blob tearing it apart; this will
happen at distances from the NS surface larger than the polar cap
size, and so the problem cannot be adequately described by the used 1D
approximation.  The case of $\jm<0$ will be addressed in a later work.
Qualitatively, however, it seems that any cascade operating as a
series of discharges would produce a population of low energy
particles, and so it should be able to adjust to any current density
in the way described above, if enough charged particles are generated.

The simulations are inconclusive about how long the thermalization
persists, because I cannot follow the plasma blob for a long time.  It
definitively works during the blob formation.  For as long as there
are low energy particles in the blob the current adjustment will work
as described above.  Note, that for current adjustment the number
density of low energy particles should be comparable to $\GJ{n}$, what
is only a very small fraction of the plasma density in the blob.  If,
however, at some time the blob runs out of low energy particles, a
macroscopic electric field will arise which can adjust the current
density by creating a separate population of backward moving particles
or/and shifting mean velocities of electrons and positrons as it is
suggested in e.g.~\citet{Scharlemann1974,Lyubarsky2009}.

I performed a full-fledged kinetic modeling of pair cascades including
all essential classes of physical processes relevant to dynamics of
electromagnetically driven cascades, listed at the beginning of
Sec.~\ref{sec:gener-numer-algorithm}.  All previous attempts to model
time-dependent cascades used on-the-spot approximation for pair
injection.  In some works fluid approximation has been used, where
electrons and positrons were represented as fluids
\citep{Levinson05,Luo/Melrose2008}.  Although the physical situation I
considered -- the Ruderman-Sutherland cascade in the polar cap -- is
different from ones studied in previous works, it is possible to
assert applicability of on-the-spot and two-fluid approximations in a
general context, based on the general picture of cascade development
inferred from my simulations.

It turns out that the delay of pair injection due to finite time
necessary for a photon to propagate before it is absorbed does not
introduce new qualitative features.  I also performed simulations
using on-the-spot approximation, when an electron-positron pair was
injected at the position and at time where and when the parent
particle emitted the pair producing photon.  Quantitatively,
on-the-spot approximation introduces error in final energies of the
relativistic particles and all depending on them cascade parameters by
a factor of several.  However, qualitatively, the results are similar,
i.e. the pattern of the plasma flow remains the same: pair formation
is quasiperiodic with plasma blobs propagating into the magnetosphere
leaving tails of modestly relativistic particles.

In the first work about modeling of time-dependent cascades by
\citet{AlBer/Krotova:1975} a zero-dimensional model was used -- only
temporal, but not spatial, variations in particle number density were
considered.  In that model the production of larger amount of
particles than necessary for screening of the electric field was due
to the time delay between the photon emissions and absorptions.  As
all later attempts to model time dependent cascades used on-the spot
approximations for pair injections
\citep{Levinson05,Beloborodov2007,Luo/Melrose2008}, the question about
importance of pair injection delay remained unanswered.  In my
simulations the overshooting in pair number density arises mainly
because of the spatial separation between the acceleration and the
pair production zones in a quite regular plasma flow.  Particles are
accelerated in the gap and must travel some distance before they can
emit high energy photons.  There are particles of only one charge sign
in the gap, and so pairs are injected at only one end of the gap.
There always exists a spatial domain with the electric field (the gap)
where pairs cannot be injected and the electric field is not regulated
directly by the pair injection.  The back reaction on the electric
field proceeds only by means of gap shrinkage, which is slow.  This
causes an overshooting in pair production and so the intermittency of
pair creation.  Inclusion of spatial and temporal delays of pair
injection due to photon propagation only exaggerates this effect, but
it does not introduce a new kind of behavior.  Hence, using
on-the-spot approximation in toy models seems to be justified.  On the
other hand, in a situation when plasma flow can become chaotic the
time delay might become a deciding factor for creation of plasma
density overshoot.

Two-fluid approximation, on the other hand, is inadequate.  The pair
plasma in the discharge zone acquires a large momentum dispersion and
some particles become mildly relativistic.  A weak fluctuating
electric field easily reverts particles of both signs and plasma
becomes essentially four-component (see
Sec.~\ref{sec:part-moment-redistr}).  In two-fluid approximation at
any given point at any time each particle specie (electrons and
positrons) can move only in one direction.  This introduces an
additional rigidity, which might be the reason why \citet{Levinson05}
got strong fluctuating electric field throughout the whole domain.
Although I considered a different physical situation and the results
described in this paper can not be directly compared with the results
of \citet{Levinson05,Luo/Melrose2008}, I think that the latter are
seriously flawed by the use of two-fluid approximation.

Now I would like to discuss how properties of cascades could manifest
in pulsar radioemission.  Pair creation is not chaotic, with clear
signatures of a limit cycle behavior; this ought to have strong
observational implications.  If, as it is widely accepted today,
pulsar radioemission is directly related to pair production, the
periodicity of cascades must be visible in power spectra of pulsar
individual pulses.  There are two characteristic time scales:
$\tau_1$, associated with the blob size, and $\tau_2$, the time
between discharges.  The size of the blob to the order of magnitude is
approximately the same for all current densities, and so $\tau_1$ is
of the same order for any reasonable current density $\jm$; $\tau_2$,
on the other hand, should be more variable.

Pulsar radioemission is highly variable on timescales comparable with
the pulsar period: emission occurs mainly in form of subpulses, in
some pulsars subpulses drift, some pulsars changes modes and/or
switches off for many periods.  This hints that current density can
fluctuate because of some processes involving the whole magnetosphere
\citep[e.g.][]{Arons1983a,Timokhin_NULLING_2009}.  Cascades can adjust
to any reasonable current density, and so the current density at a
fixed colatitude might vary on timescales much larger that
$\tau_1,\tau_2$; on the other hand, the current density varies across
the pulsar polar cap anyway.  Because of these, an individual subpulse
represents emission averaged over time and space, or, in other words,
over a range of different $\jm$'s, and so the features of cascades
along separate field lines will be smeared.  Hoverer, the time scale
associated with the size of plasma blob $\tau_1$ by the order of
magnitude remains the same and should be clearly visible in the power
spectrum.  The second time scale $\tau_2$ should be less prominent,
but, as discussed before, it might be not extremely sensitive to the
current density, and, therefore, it might manifest as a broad feature
in the power spectrum.

The blob is of the same length as the region with the accelerating
electric field.  In the Ruderman-Sutherland model this length is
small, and the corresponding timescale $\tau_1$ is less than a
microsecond.  In space charge limited flow models, on the other hands,
the length of the acceleration zone should be comparable to the NS
radius; if in this case cascades work similarly, $\tau_1$ should be of
the order of $\sim100~\mu$sec.  The second time scale, $\tau_2$,
should be substantially longer, a factor from few to hundreds.  There
are evidences of different characteristic time scales in pulsar
microstructure, from nano- to milliseconds; in some pulsars
microstructure is also quasiperiodically modulated
\citep[e.g][]{Boriakoff1976,Popov2002}.  It is not clear whether
microstructure timescales are due to polar cap cascades variability or
not, but $\tau_1,\tau_2$ can be in the range of observed
microstructure modulation times, and cascades operating as a series of
discharges should have double-timescale signature.

The problem of pulsar radioemission mechanism in notoriously difficult
and currently there is no reliable theory which could adequately
explain it.  The firmly established observational fact about pulsar
radioemission is that it is due to some collective process.  In my
simulations I saw formation of a large amplitude electrostatic wave.
Its phase velocity is larger that the speed of light, and it is not
damped via Landau damping.  In one dimension in a superstrong magnetic
field only electrostatic waves exist, but in a real pulsar such wave
can be coupled to an electromagnetic mode; if it stays superluminal,
it can escape the magnetosphere.  This wave is a collective form of
emission, as it involves coherent macroscopic plasma motion.  The
simulations are inconclusive about the fate of that wave because at
some point the numerical scheme stops resolving its wavelength; it is
also not clear how coherent the whole picture is in 3D.  May be it is
too preliminary to tell whether pulsar radioemission, or some of its
component, is related to this transient wave, but in future research
special attention should be paid to such transient waves.

\section*{Acknowledgments}

I am deeply indebted to Jonathan Arons for encouragement, support, and
innumerable exciting discussions which significantly influenced my
understanding of the problem; I am also thankful for his comments to
the draft version of the paper.  I wish to thank Yuri Lyubarsky and
Anatoly Spitkovsky for helpful discussions.  This work was supported
by NSF grant AST-0507813; NASA grants NNG06GJI08G, NNX09AU05G; and DOE
grant DE-FC02-06ER41453.

\bibliographystyle{mn2e} 

\bibliography{/home/atim/ARTICLES/Bibliographies/pulsars/pulsars_theory,/home/atim/ARTICLES/Bibliographies/pulsars/pulsars_obs,/home/atim/ARTICLES/Bibliographies/magnetars/magnetars,/home/atim/ARTICLES/Bibliographies/black_holes/black_holes,/home/atim/ARTICLES/Bibliographies/NumericalMethods/numerical_methods,/home/atim/ARTICLES/Bibliographies/Books_Physics/books_physics}

\appendix 

\section{One-dimensional time-dependent electrodynamics of the
  polar cap}
\label{sec:app_1D_electrodynamics}

\begin{figure}
  \begin{center}
    \includegraphics[width=\columnwidth]{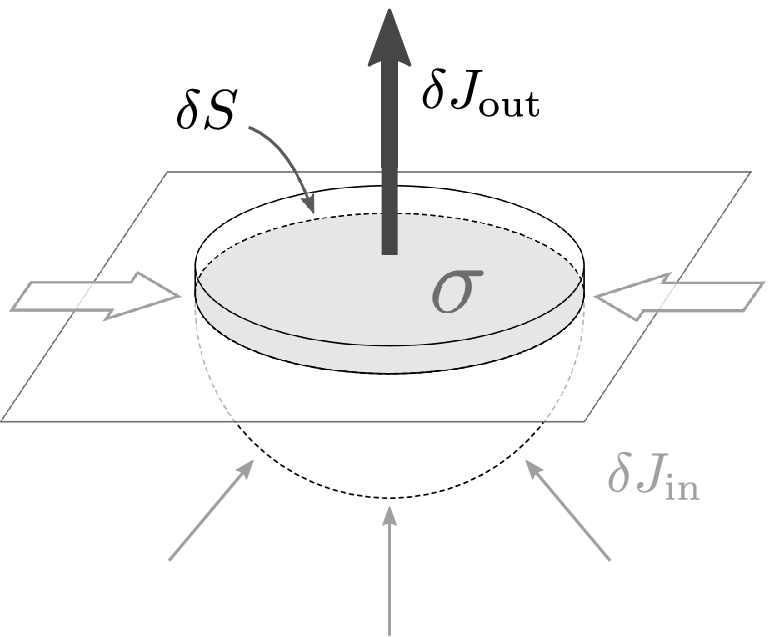}
  \end{center}
  \caption{Current flow through the surface on the NS. See text for
    explanation.}
  \label{fig:boundary_conditions}
\end{figure}

In the reference frame corotating with the NS the Gauss law is
\citep[see e.g.][]{Arons1979}
\begin{equation}
  \label{eq:gauss_law}
  \nabla\times{}E = 4\pi(\eta-\GJ{\eta})\,.
\end{equation}
In the 1D approximation the only changing component of electromagnetic
fields is the electric field parallel to the static magnetic field of
the NS.  The solution of equation~(\ref{eq:gauss_law}) is given by
\begin{equation}
  \label{eq:ex_int_rho}
  E = \AT{E}{x=0} + 4\pi\int_0^x(\eta - \GJ{\eta})\,ds \,.
\end{equation}
I am solving a non-stationary problem where boundary conditions can
change with time because the magnetosphere can response to the changes
of conditions in the polar cap.  If the electric field just outside
the NS surface $\AT{E}{x=0}$ is known at any given moment of time, the
electric field in the calculation domain can be calculated using
eq.~(\ref{eq:ex_int_rho}).  For this problem it is more convenient to
reformulate the boundary conditions on the electric field at the NS
surface $\AT{E}{x=0}$ in terms of the electric current flowing through
the system.

As $\GJ{\eta}$ does not change with time, differentiating
eq.~\ref{eq:ex_int_rho} with respect of time and using charge
conservation
\begin{equation}
  \label{eq:charge_cons}
  \DERIV{\eta}{t} + \DERIV{j}{x} = 0
\end{equation}
I get
\begin{equation}
  \label{eq:ex_e0_j}
  \DERIV{E}{t} = \AT{\DERIV{E}{t}}{x=0} - 4\pi( j - \AT{j}{x=0})\,,
\end{equation}
or
\begin{equation}
  \label{eq:ex_j}
  \DERIV{E}{t} = - 4\pi(j-\jm) \,,
\end{equation}
where 
\begin{equation}
  \label{eq:j_0_definition}
  \jm\equiv\frac1{4\pi}\AT{\DERIV{E}{t}}{x=0} + \AT{j}{x=0}\,.
\end{equation}
To clarify the meaning of $\jm$ let us consider a small region at the
NS surface, see Fig.~\ref{fig:boundary_conditions}.  NS crust can be
considered as a good conductor; the charges can accumulate only on its
surface, and the electric field in the crust is zero.  The electric
field at the NS surface $\AT{E}{x=0}=4\pi\sigma$, where $\sigma$ is
the surface charge density.  The change of the total charge in the
fiducial volume in Fig.~\ref{fig:boundary_conditions} $\delta{}q$ is
due to currents through the boundaries of the volume:
\begin{equation}
  \label{eq:delta_q}
  \delta{}q=\delta\sigma{}\delta{}S=\delta{}t(-\delta{}J_{\rm{}out}+\delta{}J_{\rm{}in})
  =\delta{}t(-\AT{j}{x=0}\delta{}S+ \delta{}J_{\rm{}in})\,.
\end{equation}
For the electric field at the NS surface I have then
\begin{equation}
  \label{eq:d_sigma_dt}
  \frac1{4\pi}\AT{\DERIV{E}{t}}{x=0} = \DERIV{\sigma}{t} = -\AT{j}{x=0} + \frac{dJ_{\rm{}in}}{dS}\,.
\end{equation}
Substituting this expression into eq.~(\ref{eq:j_0_definition}) I get
\begin{equation}
  \label{eq:j_0}
  \jm=\frac{dJ_{\rm{}in}}{dS}\,,
\end{equation}
i.e. $\jm$ is the current density which flows in the NS crust toward
the discharge zone; it causes current in the discharge zone and/or
accumulation of charges at the NS surface.  In other words, $\jm$ is
the current density that the magnetosphere wants to flow through the
cascade zone.  Eq.~(\ref{eq:ex_j}) is a convenient form for an
equation for the electric field in a problem where a large system with
very high inductivity requires some specific current density from a
much smaller system plugged into the same electrical circuit
\citep[see e.g.][]{Levinson05,Beloborodov2007}.  Note that
eq.~\ref{eq:ex_j} correctly accounts for the retardation of changes in
the electric field -- at any given point in space the electric field
changes if $j$ deviates from $\jm$; the current density $j$ is
generated by particle motion, and the latter cannot move faster than
the speed of light.

\label{lastpage}

\end{document}